\documentclass[conference]{IEEEtran}
\IEEEoverridecommandlockouts
\usepackage{cite}
\usepackage{amsmath,amssymb,amsfonts}
\usepackage{algorithmic}
\usepackage{graphicx}
\usepackage{textcomp}
\usepackage[dvipsnames]{xcolor}
\usepackage[utf8]{inputenc}
\usepackage[T1]{fontenc}  
\usepackage[normalem]{ulem}
\usepackage{amsmath}
\usepackage{bbm}
\usepackage{xcolor}

\usepackage[square,sort&compress,comma,numbers]{natbib}
\usepackage{graphicx}
\usepackage{subfigure}
\usepackage[skip=-1pt]{caption}
\usepackage{amsmath}
\usepackage{makecell}
\usepackage{amsfonts}
\usepackage{hyperref}
\usepackage{enumitem}
\usepackage{booktabs}
\usepackage{multirow}
\usepackage{mathtools}
\usepackage{multicol}
\usepackage{amsmath, bm}
\usepackage{stmaryrd}
\usepackage{longtable}
\usepackage{bbm}

\DeclareMathOperator*{\argmax}{argmax}
\usepackage[ruled,vlined,english,onelanguage]{algorithm2e}
\newcommand{\new}[1]{\textcolor{black}{#1}}

\graphicspath{ {./fig/}  }  

\usepackage{titlesec}
\def\BibTeX{{\rm B\kern-.05em{\sc i\kern-.025em b}\kern-.08em
    T\kern-.1667em\lower.7ex\hbox{E}\kern-.125emX}}
\begin{document}

\title{Intent-aware Multi-source Contrastive Alignment for Tag-enhanced Recommendation\\
}

% \author{Haolun

\author{\IEEEauthorblockN{Haolun Wu{$^{1
\dagger}$}, Yingxue Zhang{$^{2}$}, Chen Ma{$^{3}$}, Wei Guo{$^{2}$}, Ruiming Tang{$^{2}$}, Xue Liu{$^{1}$} \textit{{(Fellow of IEEE)}}, Mark Coates{$^1$}}
\IEEEauthorblockA{
1. \textit{School of Computer Science}, McGill University, $^\dagger$MILA - Quebec AI Institute. Montreal, Canada \\
haolun.wu@mail.mcgill.ca, xueliu@cs.mcgill.ca, mark.coates@mcgill.ca\\
2. \textit{Noah's Ark Lab}, Huawei.
%, Huawei Canada. Montreal, Canada \\
\{yingxue.zhang, guowei67, tangruiming\}@huawei.com\\
3. \textit{Department of Computer Science}, City University of Hong Kong. Hong Kong SAR. 
chenma@cityu.edu.hk
}}

\maketitle

\begin{abstract}
% \begin{abstract}
To offer accurate and diverse recommendation services, recent methods use auxiliary information to foster the learning process of user and item representations. Many state-of-the-art (SOTA) methods fuse different sources of information (user, item, knowledge graph, tags, etc.) into a graph and use Graph Neural Networks (GNNs) to introduce the auxiliary information through the message passing paradigm. In this work, we seek an alternative framework that is light and effective through self-supervised learning across different sources of information, particularly for the commonly accessible item tag information. We use a self-supervision signal to pair users with the auxiliary information (tags) associated with the items they have interacted with before. To achieve the pairing, we create a proxy training task. For a given item, the model predicts which is the correct pairing between the representations obtained from the users that have interacted with this item and the tags assigned to it. This design provides an efficient solution, using the auxiliary information directly to enhance the quality of user and item embeddings. User behavior in recommendation systems is driven by the complex interactions of many factors behind the users’ decision-making processes. To make the pairing process more fine-grained and avoid embedding collapse, we propose a user intent-aware self-supervised pairing process where we split the user embeddings into multiple sub-embedding vectors. Each sub-embedding vector captures a specific user intent via self-supervised alignment with a particular cluster of tags. We integrate our designed framework with various recommendation models, demonstrating its flexibility and compatibility. Through comparison with numerous SOTA methods on seven real-world datasets, we show that our method can achieve better performance while requiring less training time. This indicates the potential of applying our approach on web-scale datasets.
\end{abstract}

% \begin{IEEEkeywords}
% component, formatting, style, styling, insert
% \end{IEEEkeywords}

% \input{01-abstract}

% \title{Intent-aware Multi-source Contrastive Alignment for Tag-enhanced Recommendation}
% \author{Haolun Wu}

% \maketitle

\section{Introduction}

%===1. General background of the problem and advocate the significance of the problem.===
With the ever-growing volume of online information, users can easily access an increasingly vast number of online products and services. 
To reduce information overload and to satisfy the diverse needs of users, recommendation systems have emerged and are beginning to play an important role in modern society.
These systems are primarily interested in using the user-item interaction history to predict the users' interests and thereby recommend potential satisfactory items to users. 
However, the interaction history is usually sparse and some items may have experienced minimal interaction. This is referred to as the cold-start problem, and makes it challenging for the model to identify correct relations between users and items due to insufficient information.
In order to alleviate this problem and improve the recommendation quality, auxiliary information (e.g., tags of items, reviews of items, profiles of users) is usually introduced into the item recommendation process to enrich the modeling of user-item interactions.

%===2. A brief related work discussion to address the previous attempts to solve the problem.===
The earliest techniques for incorporating auxiliary information into recommendation primarily employ heuristics.
For instance, they directly add or concatenate the embeddings of tags with the user or item embeddings to enhance the representation ~\cite{heuristic_CantadorBV10, Linear_Fernandez-TobiasC14}. 
Later works adopt feature-based methods to encode auxiliary information into sparse feature vectors and then extract latent representations via neural networks~\cite{Rendle10_FactorizationMachine, CFA_ZuoZGJ16, DSPR_XuCLMM16}. 
Recently, motivated by the capability of Graph Neural Networks (GNNs) for exploiting higher-order interactions, many state-of-the-art (SOTA) works use GNN-based methods for modeling more complex relations between users, items, and auxiliary information~\cite{WangZWZLXG18_RippleNet, WangZXLG19_KGCN, KGAT_Wang00LC19, TGCN_ChenGTXDHW20}. 
% Representative works include: (\romannumeral1) KGAT~\cite{KGAT_Wang00LC19}, which augments the user-item graph with a knowledge graph containing fruitful auxiliary information regarding items; and (\romannumeral2) TGCN~\cite{TGCN_ChenGTXDHW20}, which builds a unified graph containing user nodes, item nodes, and tag nodes, and then employs type-aware neighbor sampling and aggregation during representation learning.
% and (\romannumeral3) TagGNN~\cite{TagGNN_MaoXZLTH20} that constructs a query-item-tag tripartite graph and takes both the item recommendation and tag completion into consideration.

%===3. Discussion of the remaining problem or the limitation that has not been solved by the prior works.===
%===4. Why the prior work/design is not able to solve the remaining issues and why the issue is difficult to solve and why it is important to solve it.===
\begin{figure}[t!]
    \centering
    \centering
    \includegraphics[width=\linewidth]{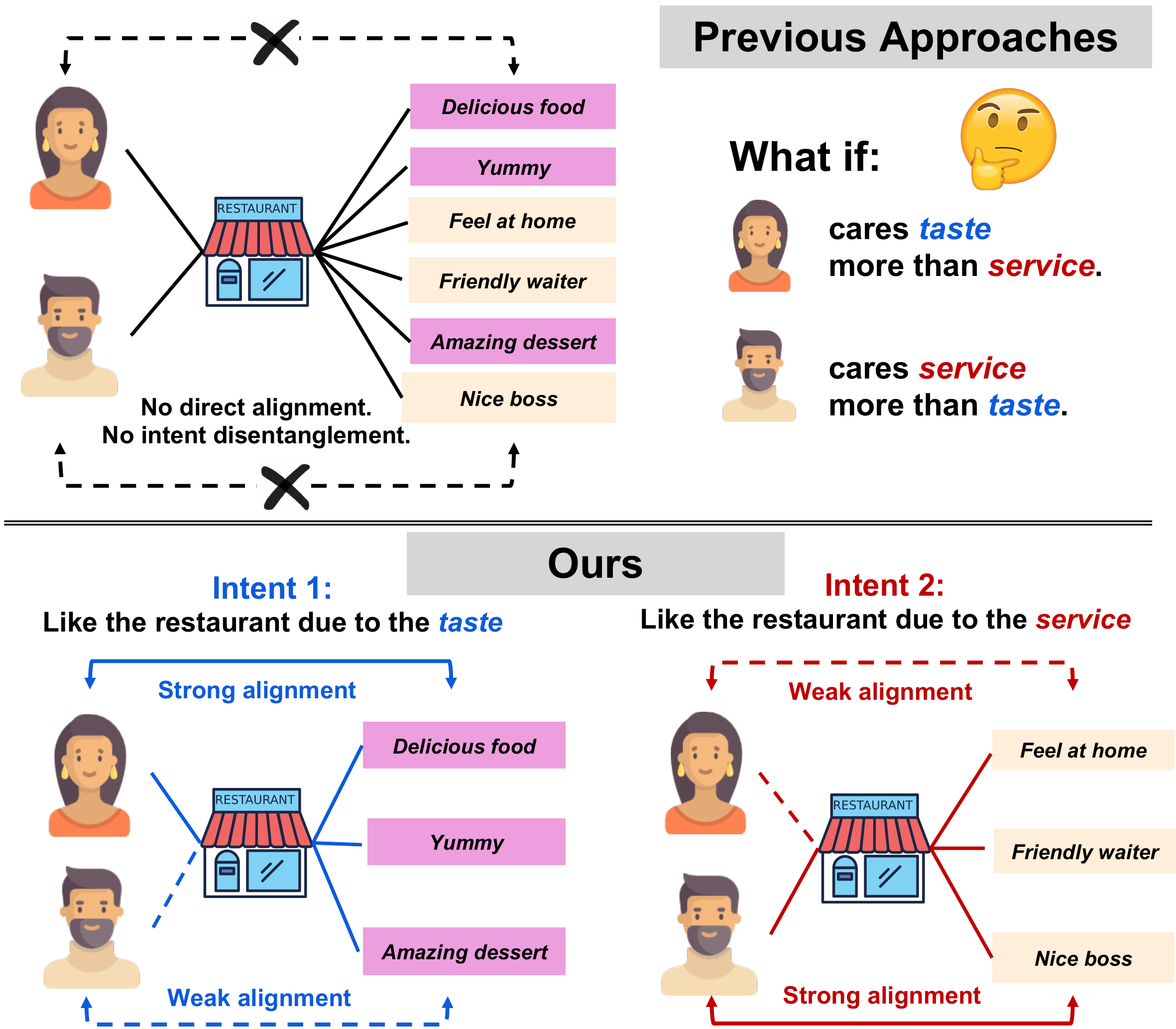}
    \vspace{-1mm}
    \caption{An illustration for showing the difference between previous approaches with ours. We not only directly align the users with tags through contrastive alignment, but also disentangle user intents during the process.}
    \label{fig:motivation}
    \vspace{-7.7mm}
\end{figure}

Despite the effectiveness of these advanced GNN-based recommendation methods with auxiliary information, we argue that they suffer from the following limitations:
\noindent\begin{itemize}[leftmargin=*]
        \item \textbf{Entanglement of User Intents.} 
    Users may have multiple intents behind their adoption of certain items~\cite{WangJZ0XC20DGCF}.
    The {\em intent} here means the motivation behind a user's interaction with an item.
    For instance, a user's intent behind visiting a restaurant may be to experience good ``service'' or to ``taste'' good food.
    Thus, ``service'' and ``taste'' are related to two kinds of user intents. Most prior approaches to employing auxiliary information in recommendation suffer from the entanglement of user intents; the models assume only one (i.e., ``interact-with'') relation between users and items~\cite{ZhangYLXM16CKE, WangZWZLXG18_RippleNet, KGAT_Wang00LC19}. Thus, the learned representations may lack interpretability.
    Further, noisy interactions (e.g., random clicks) may also provide incorrect learning guidance. 

    \item \textbf{Over-smoothing when Using Auxiliary Information.} Due to intent entanglement, prior approaches using auxiliary information~\cite{WangZWZLXG18_RippleNet,WangZXLG19_KGCN, KGAT_Wang00LC19, TGCN_ChenGTXDHW20} may further exhibit an excessive over-smoothing problem when aligning user representations with auxiliary information.
    Take the restaurant recommendation scenario as an illustration, as shown in Fig.~\ref{fig:motivation}.
    A user may like a restaurant because she likes certain attributes (tags) of it (e.g., good service, wonderful taste, etc.), and it is unnecessary for her to like all the attributes of the restaurant.
    Due to the lack of intent disentanglement, however, most previous approaches are unable to discern these details and instead match the representations of users to all of the restaurant's auxiliary information.
    Therefore, this tends to result in a severe over-smoothing issue, as the user may interact with an item due to only one or two intents.
    
    \item \textbf{Sensitivity to High-degree Nodes.} Typically, user-item interaction follows a power-law distribution~\cite{ClausetSN09_powerlaw}, with a huge number of low-degree items with very limited interaction signals. 
    This is the ``long-tail'' highlighted in~\cite{Anderson_LongTail}. 
    This phenomenon can have a major negative impact when training a GNN-based a model, since the information associated with high-degree nodes information propagates more extensively during message passing and exerts too much influence.
    The addition of new graphs (e.g., knowledge graph, item-tag graph) to the model may exacerbate this problem, as the less popular items with fewer user interactions frequently also have less auxiliary information associated with them.

\end{itemize}

In this work, we address the above limitations, focusing on tag-enhanced recommendation due to the ubiquity and accessibility of tags.
Motivated by the success of self-supervised learning for handling multi-modality data in computer vision and natural language understanding~\cite{RadfordKHRGASAM21_CLIP, yao2021filip, HjelmFLGBTB19_MIE, LarssonMS16_color, abs-1807-03748_coding}, we propose a method to efficiently bridge the collaborative filtering (CF) signal and the auxiliary semantic information. The core idea is to refine the learned representations through contrastive objectives.
In addition to modeling the user-item interaction using Bayesian Personalized Ranking (BPR)~\cite{RendleFGS2009_bpr}, we construct a self-supervised learning (SSL) task to conduct alignment between multiple sources (i.e., users, items, and tags).

To address the first limitation, we employ \textit{\textbf{Intent-aware Representation Modeling (IRM)}}.
We follow~\cite{WangJZ0XC20DGCF} to decompose user and item embeddings into multiple components, where each component captures a specific intent whose semantic meaning is identified by a corresponding tag cluster, derived using a self-supervised end-to-end clustering method.
Meanwhile, we enforce independence of different intents, ensuring that intents are effectively disentangled.

We address the second limitation by introducing an \textit{\textbf{Intent-aware Multi-source Contrastive Alignment (IMCA)}} module.
Since items interact with both users and tags, we treat them as the intermediary source.
For each intent and item, we first aggregate associated users and tags. Then, we employ contrastive learning to optimize the alignment of the aggregated representations.
We also align user and item representations. 

To address the third limitation, we propose an \textit{\textbf{Intent-aware Set-to-set Alignment (ISA)}} module to improve the performance of IMCA on cold-start users and long-tail items. For each intent (tag cluster), we identify whether two items are similar by evaluating the Jaccard index between the items' tag sets, limiting our attention to tags in the cluster. We then extend the the intent-aware contrastive alignment in IMCA to optimize the alignment of aggregated user and tag representations derived from the sets of similar items, rather than individual items. We can interpret this as an intent-aware augmentation of the user-item interactions. It leads to an augmented interaction graph with a more uniform degree distribution, mitigating the problem of high-degree nodes exerting too much influence and improving learning significantly for low-degree nodes.

We call our proposed method \textbf{I}ntent-aware \textbf{M}ulti-source \textbf{C}ontrastive \textbf{A}lignment for \textbf{T}ag-enhanced recommendation (\textbf{IMCAT}).
Due to its model agnostic property, it can be plugged into a variety of recommendation backbones.
Experimental studies on seven benchmark datasets demonstrate the effectiveness of IMCAT. It significantly improves the recommendation accuracy, especially for long-tail items. Further analysis also shows that our method requires much less time to train than other SOTA techniques.
This indicates the potential of applying IMCAT on web-scale industrial datasets.

To summarize, the contributions of this work are as follows:~\vspace{-1.5mm}
\begin{itemize}[leftmargin=*]
    \item We propose a novel method, IMCAT, for efficiently connecting the collaborative filtering signal and auxiliary tag information. 
    To the best of our knowledge, we are the first to model user intents with auxiliary information through self-supervised learning in tag-enhanced recommendation.
    \item Experiments on seven real-world datasets show that IMCAT can significantly improve the performance of the recommendation task, especially for cold-start users and long-tail items. 
    Extensive experiments on three recommendation backbones further demonstrate the compatibility of our method with different architectures.
    \item Further analysis demonstrates the superior training efficiency of our method, which strongly suggests the possibility of applying IMCAT on large industrial datasets.
\end{itemize}

\section{Related Work}
% \YX{Yingxue is done with related work} 
In this section, we summarize the related works from three perspectives in the field of recommender systems: (\romannumeral1) the use of auxiliary tags, (\romannumeral2) the use of knowledge graph, and (\romannumeral3) the use of self-supervised learning technique. 
We close this section by summarizing the difference to prior works.

\subsection{Tag-enhanced Recommendation}
As an effective and ubiquitous form of auxiliary information, tags have been widely studied in personalized recommendation systems to boost the model performance.
Earlier works use heuristic methods to combine the embeddings of tags with the item embeddings via simple addition or concatenation to achieve accuracy enhancement~\cite{heuristic_CantadorBV10, Linear_Fernandez-TobiasC14}. 
Later works adopt feature-based methods to encode tags into sparse feature vectors and then extract latent representations via neural networks~\cite{Rendle10_FactorizationMachine, JIT2R}. 
For instance, CFA~\cite{CFA_ZuoZGJ16} represents the users by the tags they have interacted with, and then uses a deep neural network to extract the  features layer-by-layer to predict the final score. 
DSPR~\cite{DSPR_XuCLMM16} makes use of a Multilayer Perceptron (MLP) to translate tag-based user and item profiles into an abstract embedding space, and then maximizes the similarity between the user representations and the relevant items. 
HDLPR~\cite{HDLPR_XuLCMM17} leverages an autoencoder to compress tag-based user and item profiles into a low-rank feature space. 
% TRSDL~\cite{TRSDL_app8050799} employs the MLP to extract item latent representations and subsequently adopts a recurrent neural network (RNN) to process consecutive historical data in order to construct user portraits. \YX{Do we really need to include the last approach \textit{TRSDL}? I don't think it relates much to tag enhanced recommendation?}

In order to better combine the information of users, items and tags  and account for the high order dependencies between them,  most recent works use Graph Neural Network based methods to further improve the model performance. For example, TGCN~\cite{TGCN_ChenGTXDHW20} builds a unified graph containing user nodes, item nodes, and tag nodes. It allows the model to leverage the contextual semantics of multi-hop neighbors in the user-tag-item graph through a message passing paradigm. Type-aware neighbor sampling and aggregation operations can learn type-specific neighborhood representations to effectively process heterogeneous node information.

\subsection{Knowledge Graph-enhanced Recommendation}
In addition to using the easily accessible tag information, prior works also investigate how to make use of knowledge graphs as auxiliary information to improve the recommendation quality.
A knowledge graph (KG) generally contains fruitful side information,  including the types of relations between entities, where the entities can be items or attributes~\cite{0002HLL18KGIntro1, 0002HLLLCD18KGINtro2}.

Several recent works have categorized KG-enhanced recommendation methods into two branches: path-based methods~\cite{YuRSGSKNH14PER, HuSZY18MCRec, WangZWZLXG18_RippleNet, WangZXLG19_KGCN, XiaHXDZYPB21KGBehavior} and embedding-based methods~\cite{ZhangYLXM16CKE, WangZXG18DKN, Xin0ZZJ19RCF, TianYRWWWL21KGNews}. 
Path-based methods extract paths from the KG that carry the high-order connectivity information and feed these paths into the predictive model. 
To handle the large number of paths between two nodes, researchers have either applied path selection algorithms to select prominent paths or defined meta-path patterns to constrain the paths.
PER~\cite{YuRSGSKNH14PER} extracts meta-path based latent features to represent the connectivity between users and items.
RippleNet~\cite{WangZWZLXG18_RippleNet} automatically discovers possible paths from an item in a user’s history to a candidate item. Embedding-based methods generally incorporate additional loss terms that capture the KG structure and use these to regularize the recommender model learning.
CKE~\cite{ZhangYLXM16CKE} uses TransR~\cite{LinLSLZ15TransR} to encode structural information among entities and an auto-encoder to capture textual and visual information. KGAT~\cite{KGAT_Wang00LC19} can be regarded as belonging to both categories: by implicitly defining the paths as the multi-hop relations between entities, it employs TransR~\cite{LinLSLZ15TransR} to learn semantic information and combines graph convolution with an attention mechanism to obtain high-quality node representations. KGIN~\cite{WangHWYL0C21_KGIN} captures user intents through the KG relations, with the goal of boosting the recommendation performance.

In this work, we focus on using tags since they are more accessible in practice. Knowledge graph information is often not available in real-world datasets.
The KG-enhanced recommendation methods can be straightforwardly adapted and used in the tag-enhanced recommendation scenario.
For applying these methods, we treat the tags and items as entities, and each connection to a specific tag as a unique relation.

\subsection{Recommendation with Self-supervised Learning}
Self-supervised learning (SSL) has been widely studied in the fields of computer vision and natural language understanding~\cite{RadfordKHRGASAM21_CLIP, yao2021filip, HjelmFLGBTB19_MIE, LarssonMS16_color, abs-1807-03748_coding}. 
Most works use the contrastive learning method~\cite{ChenK0H20_SimCLR, HjelmFLGBTB19_MI,  He0WXG20_momentum, abs-1807-03748_coding}, which maximizes the similarity of the representation of a target sample with the representations of corresponding {\em positive} samples (mutations of the target sample) and minimizes the similarity with representations of {\em negative} samples (samples known to be different). The Noise Contrastive Estimation (NCE) objective is frequently used during training. Other SSL methods proceed in a non-contrastive manner, using only positive pairs~\cite{GrillSATRBDPGAP20_BYOL, ChenH21_SimSiam}.
%Both types of SSL techniques are proven to be effective in the data augmentation and pre-training.

Nevertheless, SSL received less attention in the field of recommendation until recently. Several works aim to use SSL in the sequential recommendation context. For example, \text{S}$^3$Rec~\cite{S3Rec} uses the mutual information maximization principle to learn the correlations among attributes, items, subsequences, and sequences.
SQN~\cite{XinKAJ20_SQN} combines SSL with reinforcement learning to capture long-term user interests in sequential recommendation. 
Researchers have recently started to explore SSL techniques in Top-K recommendation.
SSLRS~\cite{YaoYCYCMHCTKE21_google_SSLRS} proposes a multi-task SSL framework for large-scale item recommendation, aiming to tackle the label sparsity problem by learning better latent relationships of item features. 
%A novel data augmentation method captures feature correlations within the proposed framework.
SGL~\cite{Wu_SGL} employs SSL to improve the accuracy and robustness of GNNs for recommendation problems without auxiliary information. KGCL~\cite{abs-2205-00976KGCL} combines SSL with KGs and proposes a cross-view contrastive learning paradigm to improve the robustness and alleviate data bias.

\textbf{Novelty and Difference.}
We summarize our novelty and differences from the following three perspectives. 
\begin{itemize}[leftmargin=*]
    \item Different from recent tag-enhanced or KG-enhanced methods that incorporate auxiliary information via GNNs, \new{we construct self-supervised objectives from multiple sources to refine the representations for collaborative filtering, naturally bringing the tag information into training. 
    This reduces the time complexity and makes it compatible with a wide variety of recommendation models.}
    
    \item Compared to recent SSL-based models in recommendation, we make the positive sampling pairing process more fine-grained to avoid embedding collapse. We propose a user intent-aware self-supervised pairing process where we split the user embeddings into multiple sub-embedding vectors. Each sub-embedding vector captures a specific user intent by self-supervised alignment with a particular cluster of tags.
    
    \item Our method is not restricted to a specific type of recommendation task (e.g., sequential recommendation), recommendation model (e.g., GNNs), or form of auxiliary knowledge (e.g., knowledge graph). The only auxiliary information we require is tags, which are usually easy to obtain. This provides our strategy with a high level of compatibility.
    
\end{itemize}

\section{Preliminaries}
\subsection{Problem Formulation}
The tag-enhanced recommendation task considered in this paper takes as input user implicit feedback and the tags associated with items as auxiliary information.
%to better depict user portraits and item features.

We denote the sets of all users, items, and tags as $\mathcal{U}$, $\mathcal{V}$, and $\mathcal{T}$, respectively. 
For each user $ u\in\mathcal{U}$, the user preference data is represented by a set of items she has interacted with as $\mathcal{I}_{u}^{+}:=\{i \in \mathcal{I}|\mathbf{Y}_{u,i}=1\}$ where $\mathbf{Y} \in \mathbb{R}^{\mathcal{|U|} \times \mathcal{|I|}}$ is the binary user-item rating matrix. 
Analogously, we use $\mathbf{Y}' \in \mathbb{R}^{\mathcal{|I|} \times \mathcal{|T|}}$ to represent the labelling history between items and tags.
We then split $\mathcal{I}_{u}^{+}$ into a training set $\mathcal{S}_{u}^{+}$ and a test set $\mathcal{T}_{u}^{+}$, requiring that  $ \mathcal{S}_{u}^{+} \cup \mathcal{T}_{u}^{+} = \mathcal{I}_{u}^{+} $ and $ \mathcal{S}_{u}^{+} \cap \mathcal{T}_{u}^{+} = \emptyset$.
% \YX{Also mention of the validation set here? or you want to say validation set will be further split from the training set}
Then the tag-enhanced top-$N$ recommendation task is formulated as: given the training item set $ \mathcal{S}_{u}^{+} $, and the non-empty test item set $ \mathcal{T}_{u}^{+}$ for user $ u $, train a model to recommend an ordered set of $N$ items $ \mathcal{X}_{u}$ such that $ \mathcal{X}_{u} \cap \mathcal{S}_{u}^{+} = \emptyset $ and $|\mathcal{X}_{u}| = N $. The model should learn from the collaborative filtering signal $\mathbf{Y}$ and the auxiliary tag information $\mathbf{Y}'$.  
The recommendation quality is evaluated by a matching metric between $\mathcal{X}_u$ and $\mathcal{T}_u^{+}$ such as Recall@$N$. 

\subsection{Bayesian Personalized Ranking}
Bayesian Personalized Ranking (BPR)~\cite{RendleFGS2009_bpr} is one of the most widely studied methods in recommendation systems for learning the user preference from the implicit user-item interaction history. 
The core idea of BPR is to maximize the ranking of an item that the user has accessed (treated as a positive sample) relative to a randomly sampled item (treated as a negative sample).
This goal is achieved via a carefully designed loss function:
% ==We denote $\Theta$ as the model parameter containing user embeddings $\mathbf{U}$ and item embeddings $\mathbf{V}$. 
\begin{align}
\mathcal{L}_{UV}&:=\mathcal{L}_{\text{BPR}}(u,v^+,v^-)=-\mathrm{log} \, \sigma \big(\hat{y}_{uv^+} - \hat{y}_{uv^-} \big) \,,
\label{eq:BPR_UV}
\end{align}
where $(u,v^+,v^-)$ is a training triplet with a positive item $v^+$ and a negative item $v^-$ for user $u$. 
$\hat{y}_{uv^+}$ refers to the relevance score between $u$ and $v^+$.

In order to better use the tag information, we adopt a similar formulation for learning the relations between items and tags. This can be viewed as recommending tags to items based on the previous item-tag pairing history.
% This is quite intuitive, as the tags assigned to an item should properly describe the item's attributes.
The loss function for this task can be formulated as:
\begin{align}
\mathcal{L}_{VT}&:=\mathcal{L}_{\text{BPR}}(v,t^+,t^-)=-\mathrm{log} \, \sigma \big(\hat{y}_{vt^+} - \hat{y}_{vt^-} \big) \,,
\label{eq:BPR_VT}
\end{align}
where $(v,t^+,t^-)$ is a training triplet with a positive tag $t^+$ and a negative tag $t^-$ for item $v$.

For simplicity, we use $\mathbf{u}, \mathbf{v}, \mathbf{t}$ to represent the individual embeddings of a user, item, and tag, respectively, in the rest of this paper.
All embeddings share the same dimension $d$.
\section{Methodology}
~\label{sec:method}
We present the technical details of the three modules comprising the proposed IMCAT method in this section. We first describe how to obtain an intent-aware representation of users and items, and how these intents connect to the auxiliary tag information. 
Then we describe the overall framework of the intent-aware contrastive alignment for efficiently combining the collaborative filtering signals and auxiliary item tagging information in a self-supervised manner.
Further, we describe how to design more diverse positive sample pairs by aligning users with the tags not only from the items they have interacted with but also with the tags from other similar items. This enriches the representations, particularly for cold start users and items.  
Finally, we elaborate upon the overall training objective that is used to optimize the whole framework.

\subsection{Intent-aware Representation Modeling (IRM)}

In this section, we discuss how to obtain intent-aware representations for users and items.
Specifically, we first explain the intent-aware embedding initialization for users and items, and then we connect each  intent sub-embedding vector with one of the semantically similar tag clusters obtained via an end-to-end clustering method.

\subsubsection{\textbf{Intent-aware Initialization for Users and Items}} 

User behavior in recommender systems is driven by the complex interactions of many factors behind the users’ decision-making processes. 
Most prior works in recommendation do not disentangle user intents. Instead, there is a single representation for each user that captures the combined effects of all intents. There is no explicit modelling of the factors that contribute to each user-item interaction.
To model the user-item interaction behavior in a more fine-grained manner, we decompose the representation of each user and item into $K$ components. Each component (sub-embedding) aims to represent a distinct user preference.
We follow \cite{WangJZ0XC20DGCF} to conduct the intent-aware initialization for users and items.
Specifically, the representations of user $u_i$ and item $v_j$ proceed as follows:
\begin{equation}
\begin{aligned}
    \mathbf{u}_i=[\mathbf{u}^1_i, \mathbf{u}^2_i, ..., \mathbf{u}^K_i] \,,\quad
    \mathbf{v}_j=[\mathbf{v}^1_j, \mathbf{v}^2_j, ..., \mathbf{v}^K_j] \,.
\end{aligned}
\end{equation}

For a fair comparison with previous methods, we keep the entire embedding size as $d$ \new{to maintain the same parameter size as other intent-unaware methods.}
Therefore, $\mathbf{u}_i, \mathbf{v}_j\in\mathbb{R}^d$ and $\mathbf{u}^k_i, \mathbf{v}^k_j\in\mathbb{R}^{d/K}$, $k=1, 2, ..., K$, where $K$ denotes the number of user intents, which is a hyper-parameter.

\subsubsection{\textbf{Self-supervised End-to-end Tag Clustering}}
We hypothesize that multiple tags with similar semantic meanings can be regarded as a common factor that biases a user to interact with items with similar traits.
For instance, consider a restaurant recommendation scenario.
A tag cluster \textit{``delicious food, yummy, amazing dessert''} may correspond to the same factor that the users like the restaurant due to the ``taste'' of the food, while a tag cluster \textit{``feel at home, friendly waiter''} can be used to explain another intent corresponding to a desire for good ``service''. 
Thus, we focus on how to cluster tags so that the $k^{th}$ tag cluster can be properly aligned with the $k^{th}$ intent embedding for users and items. 

One naive solution is to iteratively apply the K-means algorithm~\cite{Lloyd1982Kmeans} on the learned tag embeddings as the training procedure proceeds.
The tag embeddings can be trained through the objective $\mathcal{L}_{UV}+\alpha\cdot \mathcal{L}_{VT}$, where $\alpha$ is a scaling factor.
%, so that the tag embeddings can be properly learned. 
% \YX{I kind of struggle to convince what has been done can actually capture the semantic information. Usually, when we use the term semantic, it means the meaning of the text itself. If you emphasize the term "semantic", it will bias the reader to believe that you mean to use the tagging word embedding. But what you show here is the way you obtain the tag embedding is only based on u-i and i-t interaction, which is not related to semantics. So it creates a conflict. Maybe be careful when you use the term semantic.}
However, such a clustering is not optimized jointly with the downstream objective and might be sub-optimal. Instead, we employ an end-to-end self-supervised clustering approach  to adaptively obtain the tag clusters, as inspired by the papers~\cite{WangPHLJZ19AttributeCluster, Bo0SZL020StructureCluster}.
To be more precise, we directly learn tag cluster center embeddings $\bm{\mu}\in\mathbb{R}^{K\times d}$, and compute the probability of each tag belonging to each cluster. 
To achieve this, we use a Student’s t-distribution~\cite{Maaten_student} to model the probability of assigning the  tag $t_l$ to the $k^{th}$ cluster as follows:
\begin{equation}
    \mathbf{Q}_{lk}=\frac{(1+||\mathbf{t}_{l}-\bm{\mu}_k||^2/\eta)^{-\frac{\eta+1}{2}}}{\sum\limits_{i=1}^K(1+||\mathbf{t}_{l}-\bm{\mu}_{i}||^2/\eta)^{-\frac{\eta+1}{2}}}\,,
\label{eq:distribution}
\end{equation}
where $\mathbf{t}_{l}\in\mathbb{R}^d$ is the embedding of tag $t_l$, the $l^{th}$ row of $\mathbf{Q}$ is the probability of assigning tag $t_l$ to each of the $K$ clusters, and the temperature $\eta$ controls the distribution's sharpness.

\new{To build a self-supervised signal in our clustering task for an end-to-end learning}, we construct a target distribution $\mathbf{\hat{Q}}_{lk}$ which strives to push the representations closer to cluster centers, strengthening the cohesion of the clusters~\cite{Bo0SZL020StructureCluster}.
The target distribution $\mathbf{\hat{Q}}_{lk}$ is defined as follows:
\begin{equation}
    \mathbf{\hat{Q}}_{lk}=\frac{\mathbf{Q}_{lk}^2/\sum_l\mathbf{Q}_{lk}}{\sum\limits_{i=1}^K(\mathbf{Q}_{li}^2/\sum_l\mathbf{Q}_{li})} \,.
\end{equation}

We then construct a self-supervised loss objective for the end-to-end clustering as the Kullback–Leibler (KL)-divergence between the above two matrices:
\begin{equation}
    \mathcal{L}_{KL} = KL(\mathbf{\hat{Q}}||\mathbf{Q})=\sum\limits_{l=1}^{|\mathcal{T}|}\sum\limits_{k=1}^K \mathbf{\hat{Q}}_{lk}\log\frac{\mathbf{\hat{Q}}_{lk}}{\mathbf{Q}_{lk}} \,.
\end{equation}
After obtaining the matrix $\mathbf{Q}$ in Eq.~\eqref{eq:distribution}, we use a hard allocation to assign each item to one tag cluster. The assigned tag cluster index is determined by $\argmax_k(\mathbf{Q}_{lk})$ for tag $t_l$.

\subsection{Intent-aware Multi-source Contrastive Alignment (IMCA)} 

Motivated by the recent success of SSL for multi-modality data in both computer vision and natural language understanding~\cite{RadfordKHRGASAM21_CLIP, yao2021filip, HjelmFLGBTB19_MIE, LarssonMS16_color, abs-1807-03748_coding}, we introduce a new formulation by combining the two modalities of information (user-item collaborative signals $\mathbf{Y}$ and item-tag auxiliary information $\mathbf{Y'}$) into a common user-item-tag space using a contrastive learning paradigm.

Before conducting such a contrastive alignment, we need to find a bridge to connect these sources of data (i.e., users, items, and tags). 
We treat the items as a middle ground to link the information coming from the other two sources because items are present in both user-item interactions and item-tag labels.
We first explain how to use each item to construct a pair of positive samples from user embeddings and tag embeddings.  Then, we present the details of the contrastive alignment objective function employed in our framework.

\subsubsection{\textbf{Multi-source Positive Sample Construction}} 

\begin{figure}[t!]
    \centering
    \centering
    \includegraphics[width=\linewidth]{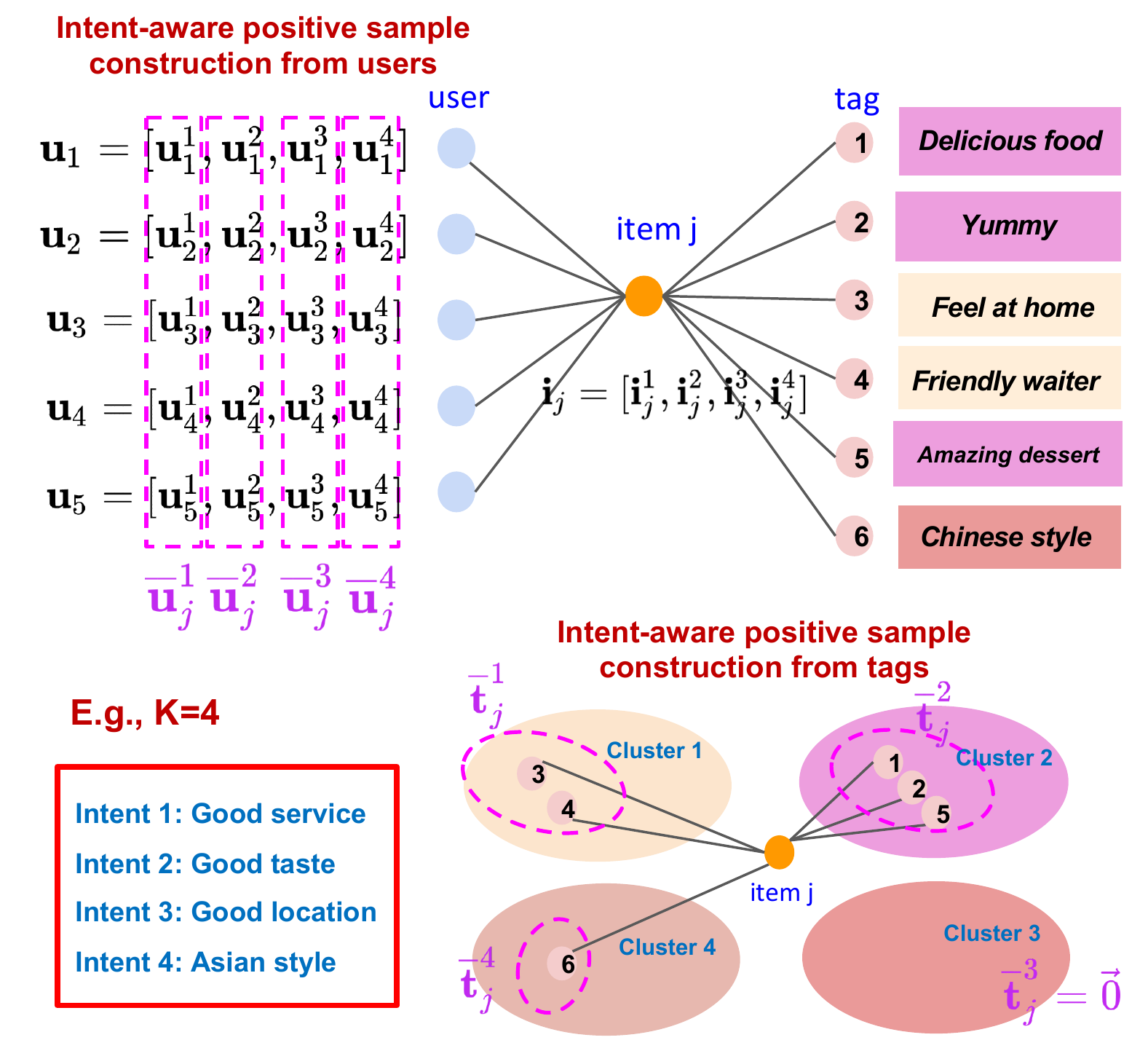}
    \caption{Intent-aware positive sample construction from multiple sources. In this restaurant recommendation example, we set the number of user intents and tag clusters ($K$) to 4.}
    \label{fig:positive_sample}
    \vspace{-6mm}
\end{figure}

We first discuss how to form the positive sample for each item from the user embeddings. 
Specifically, for a given item, we conduct an aggregation on those users who previously interacted with the item, but we design this aggregation to be intent-aware. Following the intent-aware initialization, the most intuitive way for conducting such an aggregation is to use an arithmetic average on each intent component over the user embeddings.
The positive sample representation of the $k^{th}$ intent with respect to item $v_j$ from the source of users can thus be written as:
\begin{align}
    \overline{\mathbf{u}}^k_j=\text{aggregate}(\{\mathbf{u}^k_i, \forall i\in\mathcal{I}_u(v_j)\}),
\end{align}
where $\mathcal{I}_u(v_j)$ contains the indices of all the users $u_i$ who have interacted with item $v_j$ before, i.e., $\mathbf{Y}_{u_i,v_j}=1$.
Therefore, the overall aggregation of user representations on item $v_j$ can be written as $\overline{\mathbf{u}}_j=[\overline{\mathbf{u}}^1_j, \overline{\mathbf{u}}^2_j, ..., \overline{\mathbf{u}}^K_j]\in\mathbb{R}^{d}$. 
Now we discuss how to obtain the positive samples from tag embeddings given an item.
Previously, we obtained the tag clusters through self-supervised training, where each cluster is related to a user intent.
Now, for a given item $v_j$, we compute its tag cluster embedding for each cluster through aggregating only over those tags assigned to $v_j$.
We conduct this across all $K$ clusters and all $|\mathcal{V}|$ items.
The positive sample representation of the $k^{th}$ intent with respect to item $v_j$ from the source of tags can be written as:
\vspace{-1mm}
\begin{equation}
\begin{aligned}
    \overline{\mathbf{t}}^k_j&=\text{aggregate}(\{\mathbf{t}_l, \forall l\in\mathcal{T}^k(v_j)\}),
\end{aligned}
\end{equation}
where $\mathcal{T}^k(v_j)$ contains the indices ($l$) of those tags that are associated with item $v_j$ (i.e., $\mathbf{Y}'_{v_j, t_l}=1$) and also belong to the $k^{th}$ cluster (i.e., $\argmax_{k'}(\mathbf{Q}_{lk'})=k$).
If it happens that there is no tag of item $v_j$ occurring in the $k^{th}$ cluster, we simply set $\overline{\mathbf{t}}^k_j=\vec{0}$.
To make the positive sample construction process clearer, we show an example for $K=4$ in Fig.~\ref{fig:positive_sample}.

We also observe that different items may have varying degrees of  relatedness to distinct tag clusters and user intents.
For instance, if an item has $10$ tags related to intent $1$ and only $1$ tag related to intent $2$, the item is more closely related to intent $1$ than intent $2$.
Hence, we use a vector $\mathbf{m}_j\in\mathbb{R}^K$ to store the  relatedness of $v_j$ with respect to all intents. Specifically, $\mathbf{m}_j$ is computed based on the number of $v_j$'s tags in each cluster, and the $k^{th}$ entry can be written as:
\begin{equation}
    \mathbf{m}_{j,k}=\frac{e^{|\mathcal{T}^k(v_j)|}}{\sum_{k=1}^K e^{|\mathcal{T}^k(v_j)|}}.
\end{equation}
Finally, we define $\mathbf{M}=[\mathbf{m}_{1}, \mathbf{m}_{2}, ..., \mathbf{m}_{|\mathcal{V}|}]^\intercal\in\mathbb{R}^{|\mathcal{V}|\times K}$, where each row contains the relatedness of an item to all intents.
We use this matrix for re-weighting the contrastive loss which we will introduce in the next subsection.

\subsubsection{\textbf{Contrastive Alignment for U-I-T}}
Now, we introduce our new design by combining the two information sources (user-item collaborative signals and item tagging auxiliary information) into an aligned user-item-tag space through contrastive learning. 
We aim to maximize the alignment of the pairs of positive samples from the sources of users and tags. Specifically, for the decomposed intent $k$ and item $v_j$, we aim to align the user and tag representations (i.e., $\overline{\mathbf{u}}^k_j$ and $\overline{\mathbf{t}}^k_j$).
However, one challenge here is that they have different dimensions: the positive user aggregation $\overline{\mathbf{u}}^k_j\in\mathbb{R}^{d/K}$ while the positive tag aggregation $\overline{\mathbf{t}}^k_j\in\mathbb{R}^d$.
In order to resolve this, we first use a linear layer to transform the tag aggregation to make it share the same dimension as the user aggregation.
The transformed positive tag aggregation for item $v_j$ under the $k^{th}$ intent is formulated as:
\begin{align}
    \hat{\mathbf{t}}^k_j=\mathbf{W}^k_0\cdot\overline{\mathbf{t}}^k_j+\mathbf{b}^k_0,
\end{align}
where $\mathbf{W}^k_0\in\mathbb{R}^{(d/K)\times d}$ and $\mathbf{b}^k_0\in\mathbb{R}^{d/K}$ are trainable parameters.
We can then easily compute the alignment of $\overline{\mathbf{u}}^k_j$ and $\hat{\mathbf{t}}^k_j$ through an inner-product.

Nevertheless, merely aligning $\overline{\mathbf{u}}^k_j$ and $\hat{\mathbf{t}}^k_j$ is sub-optimal, since there is no attempt to perform an intent-aware alignment of user and item representations.
To address this, we propose to merge the user-tag and user-item intent-aware alignments into a single unified alignment task: maximize the alignment between  $\overline{\mathbf{u}}^k_{j}$ (i.e., the aggregated user representation for intent $k$) and $\overline{\mathbf{z}}^k_{j'}=\hat{\mathbf{t}}^{k}_{j'}\oplus\mathbf{v}^{k}_{j'}$ (i.e., the sum of the item embedding $\mathbf{v}^{k}_{j'}$ and its corresponding aggregated tag embedding $\hat{\mathbf{t}}^{k}_{j'}$ for intent $k$) when $j=j'$, and minimize it when $j\neq j'$.
Here the ``$\oplus$'' indicates element-wise addition, so the output dimension remains the same.
In other words, we currently regard $\overline{\mathbf{u}}^k_{j}$ and $\overline{\mathbf{z}}^k_{j'}$ as a pair of positive samples only when $j=j'$.
\new{It is worthy to notice that, a direct element-wise addition may make the embeddings with larger magnitudes dominate the other.
During implementation, we normalized the $\hat{\mathbf{t}}^{k}_{j'}$ and $\mathbf{v}^{k}_{j'}$ through L2 normalization prior to the element-wise addition.}

We adopt the commonly used InfoNCE~\cite{ChenK0H20SimCLR, oord2018representation} loss formulation to maximize the cosine similarity of the correct pairings of user representations and item/tag representations in each training batch while minimizing the cosine similarity of the embeddings of the incorrect pairings. Particularly, we adopt a bidirectional contrastive alignment loss formulation~\cite{RadfordKHRGASAM21_CLIP} to ensure the pairing process across the multiple sources can be jointly exploited:
\vspace{-1mm}
\begin{align}
    \mathcal{L}_{CA}=\frac{1}{2K}\sum_{k=1}^K(\mathcal{L}^k_{u2it}+\mathcal{L}^k_{it2u}).
\label{eq:IMCA}
\end{align}

The user to item-tag ($u2it$) alignment under the $k^{th}$ intent is formalized as:
\begin{align}
    \mathcal{L}^k_{u2it}=-\sum_{j=1}^{|\mathcal{V}|}\log\frac{\exp[(\overline{\mathbf{u}}^k_{j}\circ\overline{\mathbf{z}}^k_{j})/\tau]}{\sum_{j^-=1}^{|\mathcal{N}(v_j)|}\exp[(\overline{\mathbf{u}}^k_{j}\circ\overline{\mathbf{z}}^k_{j^-})/\tau]}\cdot \mathbf{M}_{j,k},
\label{eq:u2it}
\end{align}
where ``$\circ$'' denotes the inner-product, and $\tau$ is the smoothing factor.
We use the predefined matrix $\textbf{M}$ here to capture the degree of alignment for each item with respect to each intent based on the corresponding relatedness.
$\mathbf{M}_{j,k}$ refers to the entry of $\mathbf{M}$ located at the $j^{th}$ row and the $k^{th}$ column, which denotes the relatedness of item $v_j$ to the $k^{th}$ cluster and intent. $\mathcal{N}(v_j)$ is the set of negative samples of $v_j$. 
We treat all items other than $j$ as candidate negatives.

\begin{figure*}[t!]
    \centering
    \centering
    \includegraphics[width=1\linewidth]{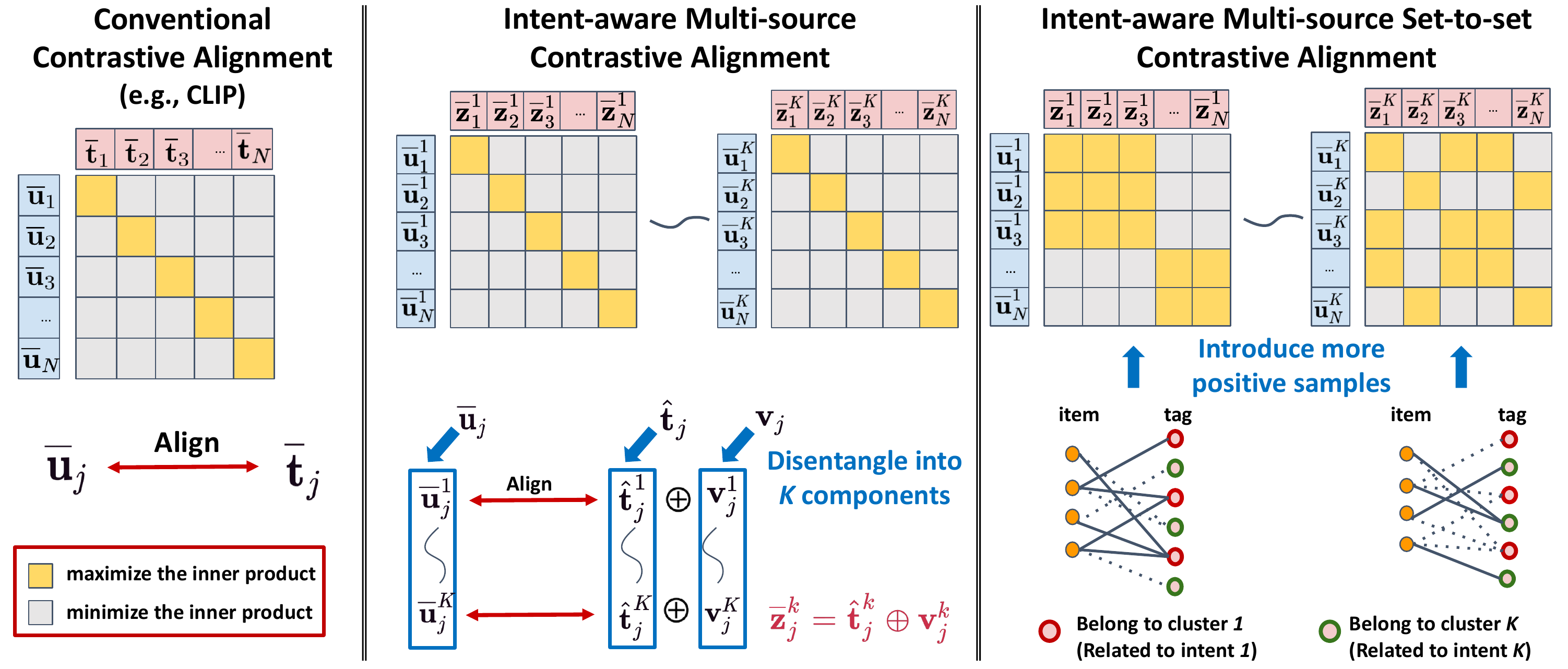}
    % \vspace{0.01mm}
    \caption{Comparison between three kinds of contrastive alignment. The first is similar to the framework adopted by CLIP~\cite{RadfordKHRGASAM21_CLIP}. The second corresponds to the mechanism proposed in the IMCA module, which models the user intents during alignment. The last is the overall mechanism of our proposed method. It boosts the performance of IMCA by introducing more positive samples to enhance the self-supervised signals, especially for those items in the long tail. For simplicity, in this figure $N=|\mathcal{V}|$.}
    \label{fig:compare}
    \vspace{-5mm}
\end{figure*}

Analogously, the item-tag to user ($it2u$) alignment under the $k^{th}$ intent can be formalized as:
\begin{align}
    \mathcal{L}^k_{it2u}=-\sum_{j=1}^{|\mathcal{V}|}\log\frac{\exp[(\overline{\mathbf{z}}^k_{j}\circ\overline{\mathbf{u}}^k_{j})/\tau]}{\sum_{j^-=1}^{|\mathcal{N}(v_j)|}\exp[(\overline{\mathbf{z}}^k_{j}\circ\overline{\mathbf{u}}^k_{j^-})/\tau]}\cdot \mathbf{M}_{j,k}.
\label{eq:it2u}
\end{align}
This alignment strategy can also be understood as maximizing the value of yellow entries and minimizing the value of grey entries in Fig.~\ref{fig:compare}.

\begin{figure*}[t]
    \centering
    \includegraphics[width=1\linewidth]{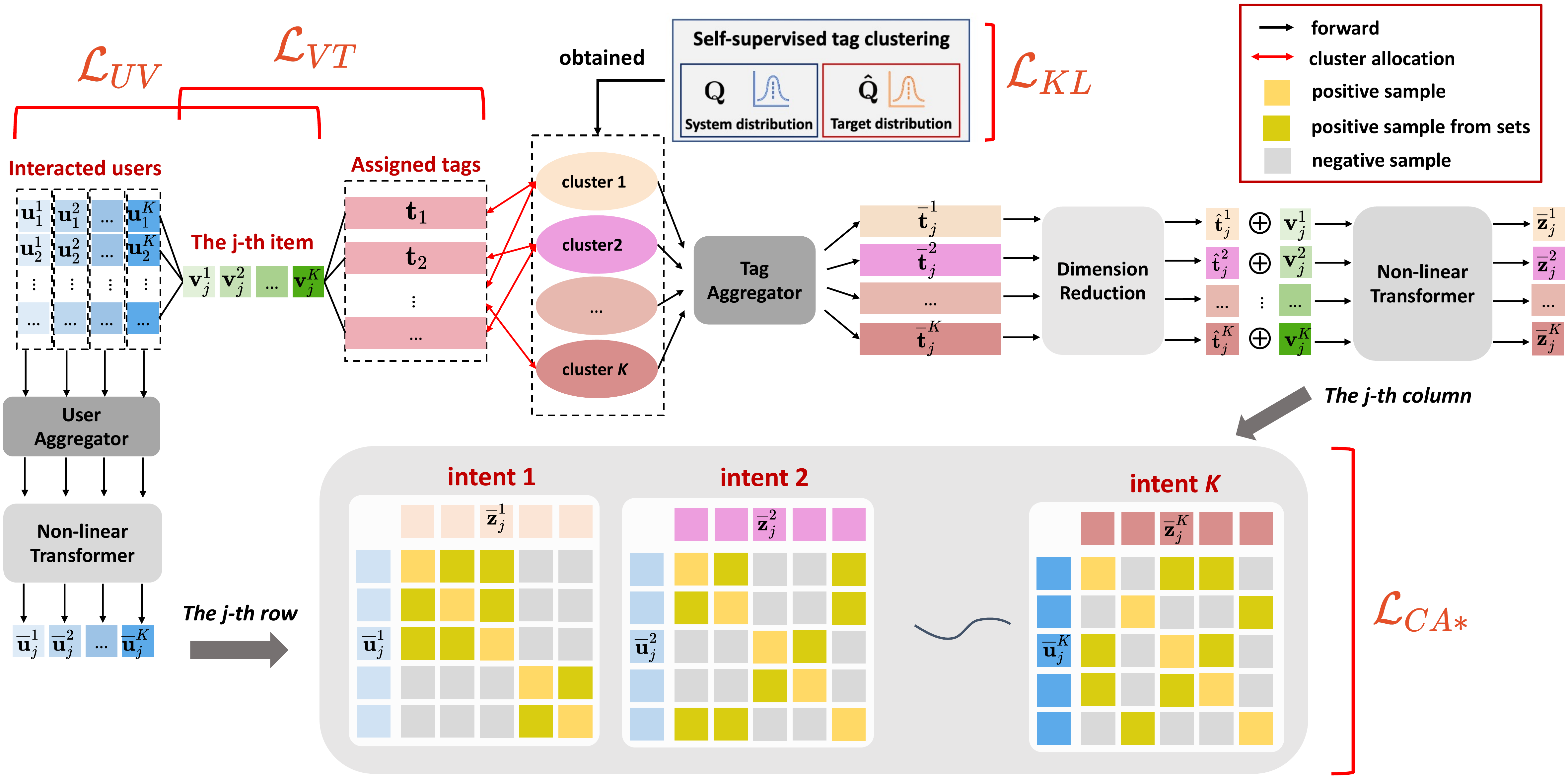}
    \caption{The overall architecture of IMCAT by taking the $j^{th}$ item as an example. Four objectives are highlighted in red color.}
    \label{fig:IMCA_overall}
    \vspace{-3.5mm}
\end{figure*}

\subsubsection{\textbf{Non-linear Transformation}}
We introduce a learnable nonlinear transformation between the representations and the contrastive loss, which further improves the quality of the learned representations. A similar strategy has been adopted for learning better representations for computer vision and NLP tasks~\cite{ChenK0H20_SimCLR}. We use a nonlinear projection head on $\overline{\mathbf{u}}^k_{j}$ and $\overline{\mathbf{z}}^k_{j}$ before the contrastive alignment in Eq.~\eqref{eq:IMCA}. Specifically, our nonlinear transformation is formalized as an MLP as follows:
\begin{equation}
\begin{aligned}
    \overline{\mathbf{u}}^k_{j}&\leftarrow \mathbf{W}^k_2\cdot\text{LeakyReLU}(\mathbf{W}^k_1\cdot \overline{\mathbf{u}}^k_{j}+\mathbf{b}^k_1),\\
    \overline{\mathbf{z}}^k_{j}&\leftarrow \mathbf{W}^k_2\cdot\text{LeakyReLU}(\mathbf{W}^k_1\cdot \overline{\mathbf{z}}^k_{j}+\mathbf{b}^k_1).
\end{aligned}
\end{equation}
Here $\mathbf{W}^k_1, \mathbf{W}^k_2\in\mathbb{R}^{(d/K)\times(d/K)}$, $\mathbf{b}^k_1\in\mathbb{R}^{d/K}$ are all trainable parameters.
The output dimension after transformation remains $d/K$.
Notice that we use a unique set of transformation parameters for each intent related representation.

\subsection{Intent-aware Set-to-set Alignment (ISA)}

In this section, we describe how to design more diverse positive sample pairs by aligning users with the tags not only from the items they have interacted with but also the tags from other similar items. This serves to enrich the representations of the cold start users and items. 

The key idea is to make use of the aggregated user representation and tag representation from other items similar to $v_j$, so as to strengthen the supervision signals on $v_j$ during the contrastive alignment process.
We compute the following similarity metric based on the Jaccard index for items $j$ and $j'$ for the $k^{th}$ intent:
\vspace{-1mm}
\begin{equation}
    s^k_{j,j'}=\frac{|\mathcal{T}^k(v_j)\cap\mathcal{T}^k(v_{j'})|}{|\mathcal{T}^k(v_j)\cup\mathcal{T}^k(v_{j'})|}.
\end{equation}
\vspace{-1mm}
Here $\mathcal{T}^k(v_j)$ is defined as before, and denotes the set of tag indices assigned to $v_j$ and also located in the $k^{th}$ tag cluster.
Then we treat any pair of items with $s^k_{j,j'}>\delta$ as similar items and regard them within the same set under the $k^{th}$ user intent factor, where $\delta$ is a pre-defined threshold.
Thus, the IMCA can be performed on all the $(\overline{\mathbf{u}}^k_j, \overline{\mathbf{z}}^k_{j'})$ pairs as long as $v_j$ and $v_{j'}$ belonging to the same set under intent $k$. This is analogous to introducing more useful positive samples in Eq.~\ref{eq:u2it} and Eq.~\ref{eq:it2u} instead of only using one positive sample to align the users and tags from the same item.

More specifically, we denote the set of items similar to item $v_j$ under intent $k$ as $\mathcal{S}^k_{j}$.
\new{Following the similar definition on both parts as that in Eq.~\ref{eq:IMCA}, we can obtain our updated loss function as:}
\vspace{-1mm}
\begin{align}
    \mathcal{L}_{CA*}=\frac{1}{2K}\sum_{k=1}^K(\mathcal{L}^{k}_{u2it*}+\mathcal{L}^{k}_{it2u*}),
\label{eq:IMCA-G}
\end{align}
\vspace{-1mm}
where $\mathcal{L}^{k}_{u2it*}$ is formulated as:
\begin{equation}
\resizebox{0.48\textwidth}{!}{
\begin{math}
    \mathcal{L}^{k}_{u2it*}=-\sum\limits_{j=1}^{|\mathcal{V}|}\frac{1}{|\mathcal{P}_j^k|}\sum\limits_{j^+=1}^{|\mathcal{P}_j^k|}\log\frac{\exp[(\overline{\mathbf{u}}^k_{j}\circ\overline{\mathbf{z}}^k_{j^+})/\tau]}{\sum_{j^-=1}^{|\mathcal{N}(v_j)|}\exp[(\overline{\mathbf{u}}^k_{j}\circ\overline{\mathbf{z}}^k_{j^-})/\tau]}\cdot \mathbf{M}_{j,k},
\end{math}}
\label{eq:u2it-G}
\end{equation}
and $\mathcal{P}_j^k$ is the set of positive samples of $v_j$ sampled from $\mathcal{S}^k_{j}$.
\new{The formulation of $\mathcal{L}^{k}_{it2u*}$ is defined analogously.}
Since the cold start items have more chance to be involved in the training process through other items sharing similar tags, the quality of the cold start item embeddings improves accordingly.
We show a comparison between different kinds of contrastive alignment methods in Fig.~\ref{fig:compare}.

\subsection{Joint Training Objective}
We adapt our model to allow forward and backward propagation for mini-batches of data. The overall training objective can be formulated as:
\begin{align}
    \mathcal{L}=\mathcal{L}_{UV}+\alpha\cdot\mathcal{L}_{VT}+\beta\cdot\mathcal{L}_{CA*}+\gamma\cdot{\mathcal{L}}_{KL},
\label{eq:overall_obj}
\end{align}
where $\alpha$, $\beta$, and $\gamma$ are scaling factors.
We show the overall architecture and the construction of these objectives in Fig.~\ref{fig:IMCA_overall}.

\section{Experiments}
\subsection{Datasets}
We evaluate our proposed method on seven real-world datasets with different domains and sparsity, where the first three datasets are all released in \textit{HetRec 2011}~\footnote{https://files.grouplens.org/datasets/hetrec2011/}~\cite{HetRec}.
The details are provided below.

\textbf{HetRec-MV} is a movie recommendation dataset.
It links movies in the MovieLens dataset with their corresponding Internet Movie Database (IMDb) web pages and Rotten Tomatoes movie reviews, where each movie is assigned with tags provided by users.

\textbf{HetRec-FM} is an artist recommendation dataset obtained from Last.fm\footnote{https://www.last.fm/}. 
It contains social networks, music tags, and music-artist listening histories of users.

\textbf{HetRec-Del} is gathered from the Delicious social bookmarking system, which contains social relations, bookmarks, and tags from users.

\textbf{CiteULike~\footnote{https://github.com/js05212/citeulike-t}} is collected from CiteULike and Google Scholar.
CiteULike allows users to create collections of articles, where abstracts, titles, and tags of each article are well-organized.
There are two versions of CiteULike datasets, CiteULike-a and CiteULike-t, both collected independently by~\cite{WangCL13CiteULike}. 
We preprocess the larger one---CiteULike-t.

\textbf{Last.fm-Tag~\footnote{http://millionsongdataset.com/lastfm/}} is collected from Last.fm, where the tracks are viewed as the items. Specifically, we take a six-month subset of this dataset by choosing the timestamps from Jan. 2015 to Jun. 2015. 
For the auxiliary information, we only keep the tags of the items.

\textbf{AMZBook-Tag} is a real-world online product recommendation dataset derived from the Amazon review datasets~\cite{HeM16AmazonData}.
We choose the part of the dataset on book recommendation as an example to study.
For the auxiliary information, we only keep the tags assigned to books.

\textbf{Yelp-Tag~\footnote{https://github.com/VC444/Yelp-Dataset-Challenge}} is adopted from the 2018 edition of the Yelp challenge. 
Here we view the local businesses such as restaurants and bars as items.
For the auxiliary information, we only maintain tags of businesses.

% For all datasets, we only maintain tags of items as the auxiliary information.
To be consistent with the implicit feedback setting, for the datasets with explicit ratings, we retain any ratings no less than four (out of five) as positive feedback and treat all other ratings as missing entries. 
We follow common practice used in prior works~\cite{sun2020_mgcf, DBLP:conf/wsdm/WangF0NC21, Ma2021_hyperknow} to filter out potential high variability users and items which have fewer than ten interactions.
Additionally, we also ensure that each tag has been allocated to at least five items. 
Table~\ref{tab:data_statistics} shows the data statistics.

\begin{table*}[t]
\centering
\caption{The statistics of datasets.}
\label{tab:data_statistics}
\resizebox{0.89\textwidth}{!}{
\begin{tabular}{lrrrrrrrr}
\toprule
& HetRec-MV & HetRec-FM & HetRec-Del & CiteULike & Last.fm-Tag & AMZBook-Tag & Yelp-Tag \\
\midrule
\#User & 2,107 & 1,026 & 1,274 & 4,011 & 18,149 & 50,022 & 39,856\\
\#Item & 3,872 & 5,817 & 5,169 & 12,408 & 14,548 & 22,370 & 26,669\\
\#Tag & 2,071 & 2,283 & 4,595 & 1,579 & 6,822 & 2,345  & 1,073\\
\midrule
\#UI & 471,482 & 57,976 & 19,951 & 94,512 & 582,791 & 731,777 & 1,009,922\\
Density & 5.78\% &  0.97\% & 0.30\% & 0.19\% & 0.22\% & 0.07\% & 0.10\%\\
\#Avg. degree & 223.77 & 56.51 & 15.66 & 23.56 & 32.11 & 14.63 & 25.34\\
\midrule
\#IT & 38,742 & 77,925 & 62,147 & 125,013 & 97,201 & 246,175 & 569,780\\
Density & 0.48\% & 0.59\% & 0.26\% & 0.64\% & 0.10\% & 0.47\% & 1.99\%\\
\#Avg. degree & 10.01 & 13.40 & 12.02 & 10.08 & 13.79 & 11.00 & 21.36\\
% \hline
% \#poor items (for PL) & 12,961 & 20,356 & 13,672 & 18,066\\
% \#rich items (for eval) & 16,058 & 24,914 & 45,162 & 45,538\\
\bottomrule
\end{tabular}
}
\vspace{-3mm}
\end{table*}

\begin{table*}[t]
\caption{The performance comparison of all methods on three backbones in terms of \textit{R@20 (Recall@20)} and \textit{N@20 (NDCG@20)} in percentage (\%). The best and the second best performing methods in each row are \textbf{boldfaced} and \underline{underlined}, respectively. The L-IMCAT method has a statistical significance for $p\leq 0.01$ compared to the best baseline method (labelled with *) based on the paired t-test.}
\resizebox{1.0\textwidth}{!}{
\vspace{1mm}
\centering
\begin{tabular}{l|cc|cc|cc|cc|cc|cc|cc}
\toprule
\multirow{2}{*}{\textbf{Model}} &
\multicolumn{2}{c|}{\textbf{HetRec-MV}} &
\multicolumn{2}{c|}{\textbf{HetRec-FM}} &
\multicolumn{2}{c|}{\textbf{HetRec-Del}} &\multicolumn{2}{c|}{\textbf{CiteULike}} & \multicolumn{2}{c|}{\textbf{Last.fm-Tag}} & \multicolumn{2}{c|}{\textbf{AMZBook-Tag}} & \multicolumn{2}{c}{\textbf{Yelp2018-Tag}} \\
& R@20 & N@20 & R@20 & N@20 & R@20 & N@20 & R@20 & N@20 & R@20 & N@20 & R@20 & N@20 & R@20 & N@20\\
\midrule
BPRMF & 13.11 & 25.74 & 16.23 & 12.92 & 17.33 & 11.83 & 16.09 & 8.97 & 33.28 & 23.45 & 14.14 & 8.12 & 8.36 & 5.41\\
NeuMF & 14.15 & 27.07 & 16.37 & 13.14 & 18.62 & 13.30 & 17.21 & 10.24 & 34.25 & 25.01 & 15.38 & 8.84 & 8.85 & 5.83\\
LightGCN & 15.09 & 29.64 & 17.01 & 13.62 & 19.85 & 15.27 & 19.14 & 11.91 & 38.73 & 29.11 & 15.89 & 9.27 & 9.37 & 6.19\\
\midrule
CFA & 14.21 & 27.34 & 16.82 & 13.44 & 18.68 & 13.42 & 17.31 & 10.64 & 34.23 & 24.93 & 15.14 & 8.65 & 8.82 & 5.81\\
DSPR & 14.62 & 28.32 & 16.94 & 13.51 & 18.32 & 13.13 & 17.42 & 10.77 & 35.30 & 26.22 & 15.39 & 8.87 & 8.84 & 5.86\\
TGCN & 15.29 & 29.84 & 19.22 & 15.31 & 20.16 & 15.74 & 21.06 & 12.71 & 43.13 & 31.62 & 17.09 & 9.96 & 9.76 & 6.47\\
\midrule
CKE & 14.28 & 27.61 & 16.78 & 13.20 & 18.76 & 13.60 & 19.18 & 11.94 & 38.21 & 28.03 & 16.54 & 9.42 & 9.09 & 6.02\\
RippleNet & 14.78 & 28.69 & 16.92 & 13.47 & 18.93 & 13.67 & 19.81 & 12.37 & 39.55 & 29.12 & 16.67 & 9.54 & 9.32 & 6.18\\
KGAT & 14.99 & 28.93 & 17.34 & 14.18 & 19.31 & 14.72 & 20.09 & 12.48 & 40.23 & 29.63 & 16.79 & 9.61 & 9.39 & 6.23\\
KGIN & 15.30 & 29.98 & 20.01 & 15.87 & 20.13 & 15.67 & 22.03$^*$ & 13.08$^*$ & \underline{44.23}$^*$ & \underline{32.72}$^*$ & 16.81 & 9.63 & 9.97 & 6.67\\
\midrule
SGL & 15.03 & 29.11 & 19.44 & 15.57 & 19.58 & 14.96 & 20.74 & 12.59 & 43.18 & 31.75 & 16.92 & 9.88 & 9.85 & 6.53\\
KGCL & \underline{15.42}$^*$ & \underline{30.24}$^*$ & 20.55$^*$ & 16.08$^*$ & \underline{20.23}$^*$ & \underline{15.82}$^*$ & 21.41 & 12.90 & 43.62 & 31.95 & 17.12$^*$ & 10.01$^*$ & 10.00$^*$ & 6.69$^*$\\
\midrule
B-IMCAT & 15.13 & 29.31 & 17.86 & 14.50 & 19.94 & 15.42 & 19.24 & 12.13 & 40.27 & 29.74 & 15.99 & 9.39 & 9.39 & 6.25\\
N-IMCAT & 15.32 & 30.16 & \underline{20.76} & \underline{16.26} & 20.15 & 15.72 & \underline{22.15} & \underline{13.14} & 44.01 & 32.31 & \underline{17.21} & \underline{10.04} & \underline{10.04} & \underline{6.72}\\
L-IMCAT & \textbf{16.22} & \textbf{33.52} & \textbf{21.25} & \textbf{17.09} & \textbf{21.58} & \textbf{16.82} & \textbf{22.87} & \textbf{13.59} & \textbf{46.73} & \textbf{33.61} & \textbf{17.72} & \textbf{10.51} & \textbf{10.41} & \textbf{6.94}\\
\bottomrule
\end{tabular}}
\label{tab:overall_performance}
\vspace{-5mm}
\end{table*}

\subsection{Evaluation Protocols}
We split user-item interactions into the training set, validation set, and testing set based on the ratios 7:1:2. 
The performance of all models is evaluated in terms of Recall@$N$ and NDCG@$N$. 
Recall@$N$ indicates the coverage of true items that appear in the top-$N$ recommended items. 
NDCG@$N$ (normalized discounted cumulative gain) is a measure of ranking quality~\cite{JarvelinK02ndcg}. 
All experiments are run 5 times with the same data partition in each trial but different parameter initializations.
The average results are reported in Table~\ref{tab:overall_performance}.

\subsection{Methods Studied}
% The aim of this work is not to propose a novel recommendation model from scratch, but to propose an efficient method on aligning multi-modality data sources to make full use of the collaborative signals and auxiliary information.
To show the compatibility of our method, we apply IMCAT on three recommendation backbones. \textbf{BPRMF}~\cite{KorenBV09MF} is MF-based, \textbf{NeuMF}~\cite{HeLZNHC17NeuMF} is MLP-based, and \textbf{LightGCN}~\cite{he2020_lgcn} is GNN-based.
The reason we choose them is because the MF-based model is the most widely used in recommendation and has been demonstrated to be efficient in real-world cases.
The MLP-based method is a generalized version of the ML-based method, which can capture more complex relations between users and items.
The GNN-based model is the state-of-the-art, and can achieve competitive performance with all current techniques.
For simplicity, we denote the models by applying the above three backbones with our IMCAT as \textbf{B-IMCAT}, \textbf{N-IMCAT}, and \textbf{L-IMCAT}, respectively.

In addition to the choice of recommendation backbones, we compare our method with the following approaches from four categories:

\textit{\textbf{Methods without Auxiliary Information}}:\vspace{-0.4mm}
\begin{itemize}
    \item \textbf{BPRMF}~\cite{KorenBV09MF} is one of the most widely used recommendation baselines and adopts a pair-wise ranking loss.
    \item \textbf{NeuMF}~\cite{HeLZNHC17NeuMF} is a generalized version of MF-based methods and uses an MLP architecture to capture non-linear feature interactions.
    \item \textbf{LightGCN}~\cite{he2020_lgcn} is one of the SOTA GNN-based methods and can capture higher-order relations between users and items using a GNN.
\end{itemize}

\textit{\textbf{Tag-enhanced Methods}}:\vspace{-0.4mm}
\begin{itemize}
    \item \textbf{CFA}~\cite{CFA_ZuoZGJ16} uses a sparse autoencoder to obtain latent representations of user profiles, on which user-based CF is applied for recommendation.
    \item \textbf{DSPR}~\cite{DSPR_XuCLMM16} leverages MLPs with shared parameters to process tag-based features for extracting user and item representations.
    \item \textbf{TGCN}~\cite{TGCN_ChenGTXDHW20} targets on tag-enhanced recommendation. 
    It employs type-aware neighbor sampling and aggregation on a unified graph.
    % \item \textbf{TagGNN}~\cite{TagGNN_MaoXZLTH20}. This method also focuses on the tag-enhanced recommendation specifically.
    % It utilizes heterogeneous graph neural networks with multiple types of nodes and edge on a query-item-tag graph.
    
\end{itemize}

\textit{\textbf{KG-enhanced Methods}}:\vspace{-0.4mm}
\begin{itemize}
\item \textbf{CKE}~\cite{ZhangYLXM16CKE} combines collaborative filtering with semantic information about items by employing TransR~\cite{LinLSLZ15TransR}.
\item \textbf{RippleNet}~\cite{WangZWZLXG18_RippleNet} propagates user preference over the knowledge graph via constructed paths rooted at each user. 
It enriches user representations by combining them with nearby item representations.
\item \textbf{KGAT}~\cite{KGAT_Wang00LC19} is one of the SOTA KG-based methods. It performs attention-aware graph convolution over the knowledge graph to capture high-order connectivity.
\item \textbf{KGIN}~\cite{WangHWYL0C21_KGIN} aims to extract latent user intents in the knowledge graph. 
It performs a relational path-aware aggregation for both the intent-aware collaborative filtering graph and the knowledge graph.
    
\end{itemize}

\textit{\textbf{SSL-based Methods}}:\vspace{-0.4mm}
\begin{itemize}
    \item \textbf{SGL}~\cite{Wu_SGL} improves the graph-based CF framework with augmented structure-based self-supervised signals.
    \item \textbf{KGCL}~\cite{abs-2205-00976KGCL} proposes a cross-view contrastive learning paradigm by constructing self-supervised signals from the structure of both the CF graph and the knowledge graph.
\end{itemize}

\subsection{Implementation Details}
We optimize all models using the Adam optimizer with the Xavier initialization procedure~\cite{GlorotB10Xavier_init}. 
We fix the embedding size to 64 and the batch size to 1024 for all methods. 
\new{We additionally fix the learning rate and weight decay both to $1e^{-3}$, and smoothing factors $\eta$ and $\tau$ both to $1$ in all experiments.}
When constructing the ranking loss objective, every positive item is associated with one sampled negative item for an efficient computation. 
Grid search is applied to choose the scaling factors $\alpha, \beta, \gamma$ are tuned from $\{1e^{-3}, 1e^{-2}, 1e^{-1}, 1, 5, 10\}$. 
The threshold factor $\delta$ in the ISA module is selected from $\{0.1, 0.3, 0.5, 0.7, 0.9\}$.
% For the base models NeuMF, MGCCF, and LightGCN, we use the author-provided implementations. 
% We use MGCCF and LightGCN with one graph convolution layer. 
The number of tag clusters (the number of intents) is selected from $\{1, 2, 4, 8, 16\}$.
Since the embedding from the early training stage is less informative, we first pre-train the model without activating the clustering loss for obtaining meaningful tag embeddings.
(We set the number of the pre-training epochs to 500.)
Then, the end-to-end clustering technique is applied to obtain the cluster membership of each tag. 
To avoid instability during training, the cluster membership of each tag is updated every 10 iterations during training.
In order to further ensure the intents are fully disentangled, we encourage independence of different intents by minimizing their correlation following the approach in~\cite{WangHWYL0C21_KGIN}.
For all baseline methods, we use the author-provided implementations. 
We use two graph convolution layers for all GNN-based methods. 
The total number of training epochs is set to 3,000 for all models with an early stopping design, i.e., premature stopping if Recall@20 on the validation data does not increase for 100 successive epochs.
All experiments are conducted on GPU machines (Nvidia Tesla V100 32G).
% \vspace{-4mm}

\subsection{Overall Performance Comparison}
The overall performance comparison is shown in Table~\ref{tab:overall_performance}.\\ 
\textbf{Observations about our method.} 
(\romannumeral1) Comparing with all the baselines that strive to make effective use of auxiliary information, our proposed IMCAT method with the LightGCN backbone achieves the best performance for all evaluation metrics. 
The superiority is significant and consistent across all seven datasets for three different backbone models. 
(\romannumeral2) L-IMCAT consistently outperforms N-IMCAT and B-IMCAT. 
This is reasonable due to the fact that LightGCN is a much stronger base model compared to NeuMF and BPRMF.
This also indicates that the GNN can capture more complex relations between nodes.
(\romannumeral3) Compared to SOTA tag-enhanced and KG-enhanced methods such as TGCN and KGIN, L-IMCAT achieves a better performance due to the constructed self-supervised signals from multiple sources for enhancing the learning process.
As such, the collaborative filtering signals and auxiliary tag information are efficiently fused.
(\romannumeral4) Compared to a SOTA SSL-based method such as KGCL, L-IMCAT achieves better performance because the contrastive objective is constructed not merely based on the topological information, but also takes into account the user intents.
(\romannumeral5) In comparison to previous SOTA GNN-based methods, our proposed IMCAT can achieve highly competitive performance even when applied on a simple base model that does not employ GNN techniques (e.g., NeuMF + IMCAT).
This strongly suggests the effectiveness of our intent-aware multi-source contrastive alignment.

\textbf{Other observations.}
(\romannumeral1) The methods lying in the first block generally do not achieve good performance due to not using auxiliary information.
(\romannumeral2) For tag-enhanced methods, CFA and DSPR do not perform well since both of them use tag-based user profiles to learn user preferences. 
Since we do not know which user provided each tag, for a specific user $u$, we can only use all tags assigned to items that $u$ interacted with to construct the user profile.
This is sub-optimal since a user does not necessarily like all the tags assigned to the items she interacts with.
TGCN achieves good performance by constructing a unified graph containing user-item-tag nodes and proposes a type-aware neighborhood aggregation method to fuse multi-source information.
(\romannumeral3) For KG-enhanced methods, KGAT is better than CKE and RippleNet since it combines the advantages of both methods and further boosts the performance by adopting the attention mechanism during message passing.
KGIN is superior to KGAT since it models user intents during training.
(\romannumeral4) The SSL-based methods are very strong baselines since they introduce more supervision signals through SSL techniques.
This significantly improves the learning process for low-degree nodes.
KGCL is even stronger than SGL because KGCL constructs self-supervised tasks by capturing the topological information from both the CF graph and the knowledge graph, while SGL only focuses on the CF graph.

\subsection{Ablation Analysis and Case Study}
Due to the space limitation, we use the NeuMF and LightGCN as the backbone models to display the ablation analysis unless otherwise specified.
% We choose NeuMF because it does not require expensive message passing in those GNN-based methods, but can still reach competitive performance when applied with our proposed IMCAT framework, compared to other SOTA methods. 
% We choose LightGCN because it is the strongest baseline model.
% We choose three datasets for showing the results: HetRec-Del, CiteULike, and Yelp-Tag.
For the following analysis, we observe a similar trend across other recommendation models and datasets we studied.
\new{The first three analysis corresponds to the ablation study on the three components proposed in Section~\ref{sec:method}, respectively.}

\subsubsection{\textbf{Analysis on the Number of Intents}}
\begin{figure}[t!]
    \centering
    \centering
    \includegraphics[width=\linewidth]{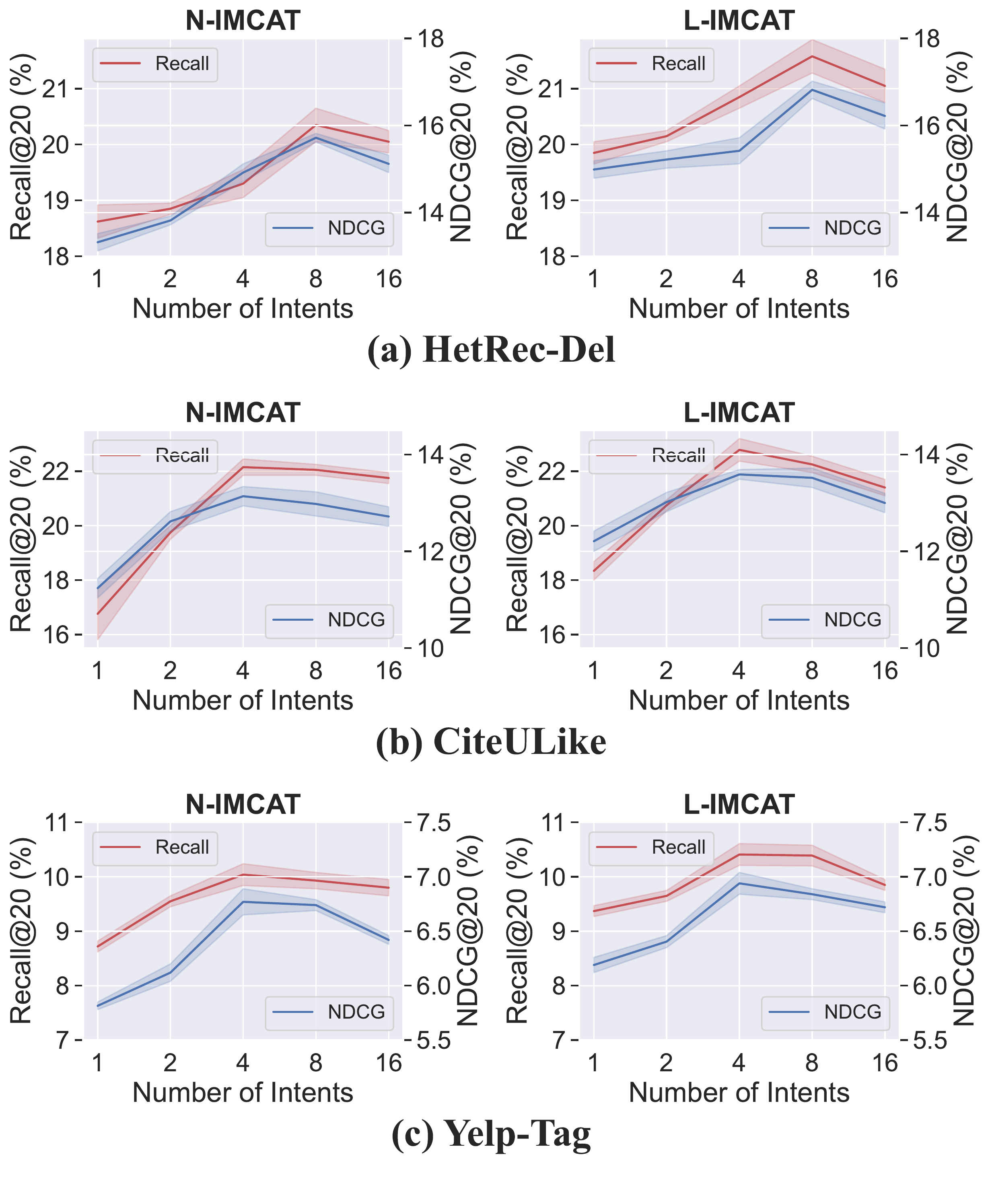}
    \vspace{-7mm}
    \caption{Impact of the number of intents ($K$). The first column shows the results when using NeuMF as the backbone model, while the second column corresponds to using LightGCN.}
    \label{fig:intent}
    \vspace{-6mm}
\end{figure}

We study the influence of
the number of intents ($K$).
We vary $K$ in the range $\{1, 2, 4, 8, 16\}$ and illustrate how the performance changes with two different models in \new{Fig.~\ref{fig:intent}}. 
The first column indicates using NeuMF with IMCAT, while the second column refers to using LightGCN with IMCAT.

As illustrated, we observe that:
(\romannumeral1) A small $K$ cannot lead to a good performance since the user intents are not sufficiently disentangled. 
For instance, when $K=1$, (this is equivalent to entangling all user intents together), we cannot distinguish the real reason why a user clicks an item. 
Thus, it is hard to capture users' real preferences.
(\romannumeral2) Generally speaking, increasing $K$ can improve the performance. 
This strongly indicates the effectiveness of the intent-aware modelling.
However, the performance drops for large $K$.
One reason may be that the intents become too fine-grained to carry useful information.
Another reason may be that the dimension size for each intent component becomes too small (i.e., $d/K$).
(\romannumeral3) Generally, $K=4$ and $K=8$ lead to the best performance.
We also observe a larger $K$ is required for HetRec-Del to achieve the optimal outcome compared to the other two datasets.
One possible reason is that HetRec-Del has more tags: it has three or four times as many tags as the other two datasets. 
As a result, HetRec-Del may contain more kinds of user intents.

\subsubsection{\textbf{Analysis on Designs for Multi-source Alignment}}

We study the impact of each design for the multi-source alignment (in the IMCA module).
Specifically, our proposed method is to align $\overline{\mathbf{u}}^k_{j}$ (user, short for U) and $\overline{\mathbf{z}}^k_{j}$ (item with tag, short for IT) for item $v_j$ under the $k^{th}$ intent.
To verify whether this multi-source alignment among U-I-T is necessary, we first study what the performance will be if we \underline{remove} some alignments. Thus: (\romannumeral1) without (w/o) UIT $\Rightarrow$ \textit{no alignment at all},  (\romannumeral2) w/o UI $\Rightarrow$ \textit{only align U and T}, and (\romannumeral3) w/o UT $\Rightarrow$ \textit{only align U and I}. 
Later, we investigate the influence of the non-linear transformation (NLT) layer before the alignment.
% For the IRM, we study the effect of intent independence modeling (short for IND).

The results are shown in Table~\ref{tab:component}. 
Overall, we observe that removing any design in our original method impairs the performance.
This indicates that each design we studied is useful.
In more detail, ``w/o UIT'' leads to extreme performance degradation. 
This is not surprising since this is equivalent to removing the alignment mechanism entirely, so that the model cannot capture any relation among the multiple sources.
Additionally, ``w/o UI'' outperforms ``w/o UT'' in all scenarios. 
The reason may be that the relation between U and I can be captured in other loss terms (e.g., $\mathcal{L}_{UV}$), whereas the relation between U and T can only be modelled through the alignment. 
Thus, the alignment between U and T is more important than that between U and I.
The performance of ``w/o NLT'' indicates that applying the non-linear transformation before the alignment is effective.
This finding is similar to that in~\cite{ChenK0H20SimCLR}.
% Lastly, the performance of ``w/o IND'' demonstrates that the independence modeling of intents is crucial.

% Otherwise, the intent-aware modelling in the whole framework cannot show its capability on capturing users' real preference.

\begin{table}[t]
\caption{Analysis on the effect of each design in the IMCA module. ``w/o'' refers to ``without''.}
\resizebox{0.5\textwidth}{!}{
\centering
\begin{tabular}{l|cc|cc|cc}
\toprule
\multirow{2}{*}{Model} & 
\multicolumn{2}{c|}{\textbf{HetRec-Del}} &
\multicolumn{2}{c|}{\textbf{CiteULike}} &
\multicolumn{2}{c}{\textbf{Yelp-Tag}}\\
& Recall & NDCG & Recall & NDCG & Recall & NDCG\\
\midrule
\textbf{N-IMCAT} & \textbf{20.15} & \textbf{15.72} & \textbf{22.15} & \textbf{13.14} & \textbf{10.04} & \textbf{6.72}\\
N-IMCAT w/o UIT & 18.97 & 14.00 & 17.82 & 11.94 & 8.88 & 5.94\\
N-IMCAT w/o UT & 19.17 & 14.09 & 19.72 & 12.41 & 9.26 & 6.11\\
N-IMCAT w/o UI & 19.82 & 15.14 & 21.67 & 12.98 & 9.78 & 6.44 \\
N-IMCAT w/o NLT & 20.08 & 15.60 & 22.08 & 13.06 & 9.82 & 6.49\\
% N-IMCAT w/o IND & 19.18 & 14.04 & 20.01 & 12.56 & 9.21 & 6.10\\
\midrule
\textbf{L-IMCAT} & \textbf{21.58} & \textbf{16.82} & \textbf{22.87} & \textbf{13.59} & \textbf{10.41} & \textbf{6.94}\\
L-IMCAT w/o UIT & 19.91 & 15.03 & 19.84 & 12.40 & 9.45 & 6.27\\
L-IMCAT w/o UT & 19.99 & 15.12 & 20.04 & 12.45 & 9.76 & 6.41\\
L-IMCAT w/o UI & 20.03 & 15.36 & 21.79 & 13.04 & 9.89 & 6.57 \\
L-IMCAT w/o NLT & 20.26 & 15.87 & 22.14 & 13.12 & 10.02 & 6.70\\
% L-IMCAT w/o IND & 20.07 & 15.58 & 20.78 & 12.62 & 9.82 & 6.49\\
\bottomrule
\end{tabular}
}
\label{tab:component}
\vspace{-5mm}
\end{table}

\subsubsection{\textbf{Analysis on the Threshold for Set-to-set Alignment}}
We analyze the effect of the threshold $\delta$ for the intent-aware item grouping with respect to the recommendation performance.
We tune $\delta$ from $\{0.1, 0.3, 0.5, 0.7, 0.9\}$ to show the trend of change in performance (i.e., Recall@20).
In order to illustrate the results more clearly, for a given model and dataset, we first compute the performance by removing the ISA module.
Then, for each setting, we report the proportion its performance reaches with respect to the result without using the ISA module.
A value larger than one means the performance improves, while a value smaller than one means the performance deteriorates.

\begin{figure}[t!]
    \centering
    \centering
    \includegraphics[width=\linewidth]{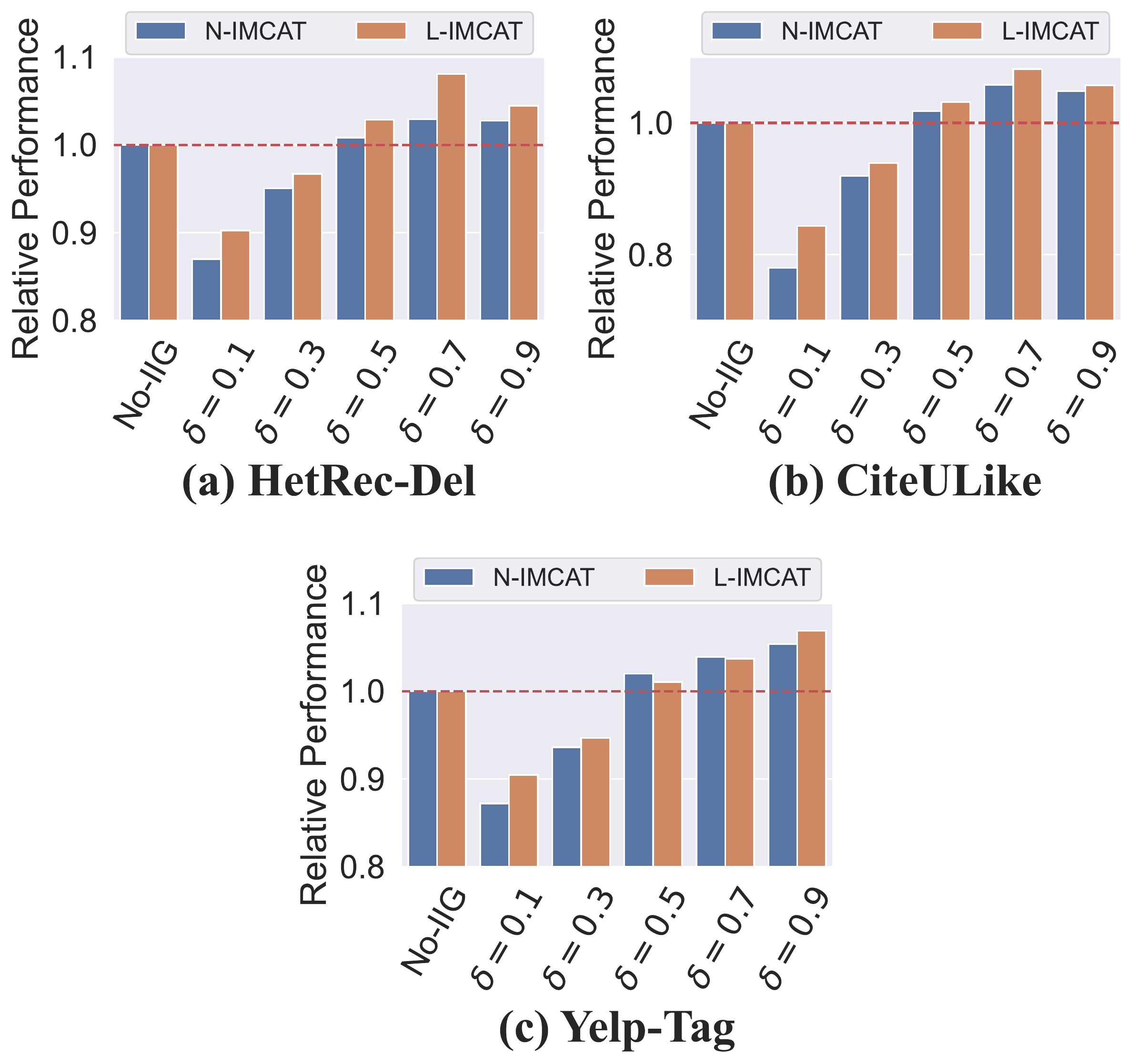}
    \caption{Analysis on the threshold for intent-aware set-to-set alignment. We present how much proportion that each setting's performance can reach compared to the result produced without this set-to-set alignment module.}
    \label{fig:threshold}
    \vspace{-5mm}
\end{figure}

As illustrated in Fig.~\ref{fig:threshold}, we first observe that the performance is even worse than not using the ISA module when $\delta=0.1$ and $\delta=0.3$.
This is because a very small threshold may introduce far too many items, including some that are not very similar.
Conducting alignment between the user aggregation and tag aggregation may then yield misleading supervision signals, which is harmful for learning the users' real preference.
When $\delta\geq0.5$, we observe the effectiveness of the set-to-set alignment as the relative performance is larger than one.
Overall, $\delta=0.7$ and $\delta=0.9$ can achieve the best performance since the items in the same set selected under these thresholds are the most similar based on the item-tag assignment matrix.

\subsubsection{\textbf{Benefits for Long-tail Items}}
As mentioned earlier, GNN-based recommendation models generally suffer from the long-tail issue and are prone to high-degree nodes.
To verify whether IMCAT is able to alleviate this problem, we follow the approach in~\cite{abs-2205-00976KGCL} to split all items into 5 groups based on the popularity. Each group contains the same number of items.
The number of interactions increases from group 1 ($G_1$) to group 5 ($G_5$).
Then we report each group's contribution to the overall Recall@20 as in~\cite{Wu_SGL}. 
For a better presentation, we normalize the performance in each group into the range $[0,1]$ by dividing by the performance obtained by the best model for that group.

\begin{figure}[t!]
    \centering
    \centering
    \includegraphics[width=\linewidth]{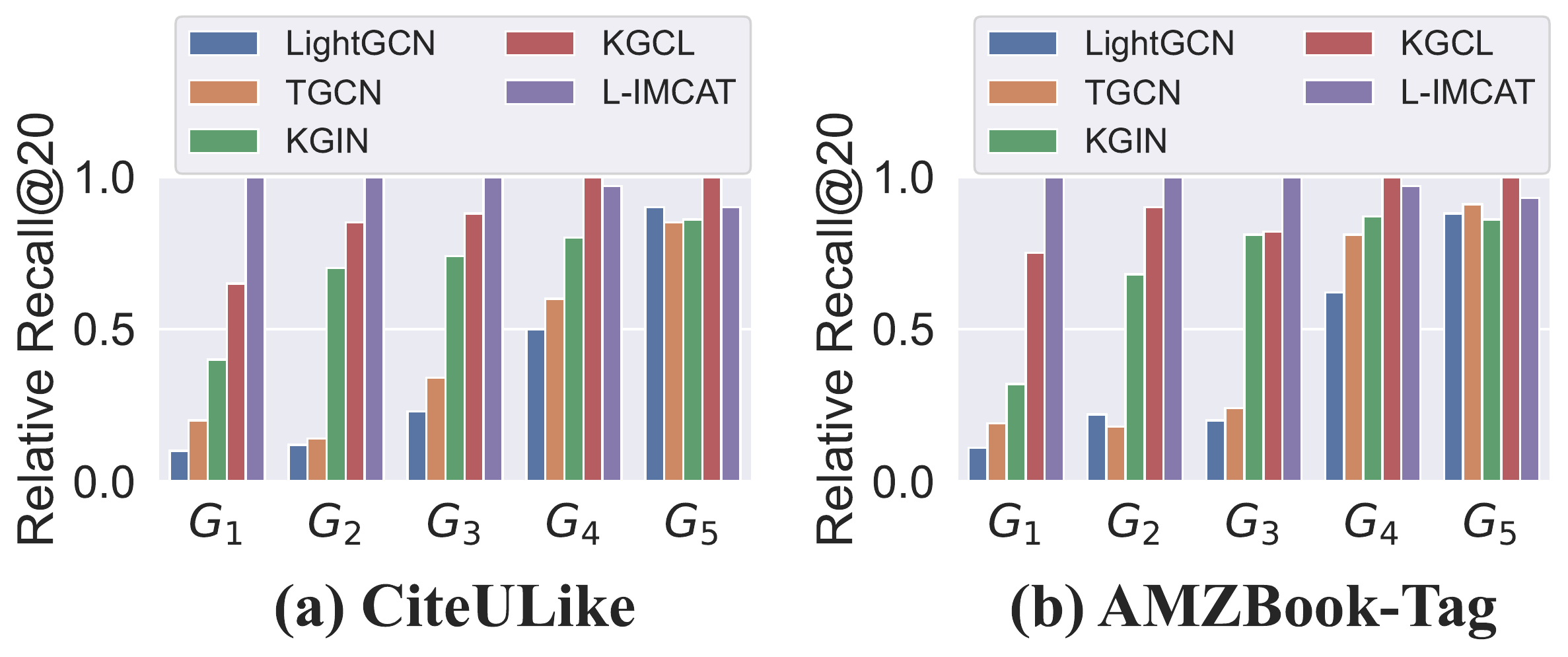}
    \caption{Performance comparison on groups of items with different interaction degrees. Performance is normalized into range $[0, 1]$ for each group.}
    \label{fig:long-tail}
    \vspace{-6.5mm}
\end{figure}

We choose the best GNN-based model from each category for this comparison.
As illustrated in Fig.~\ref{fig:long-tail}, a pure GNN-based model that only focuses on the collaborative filtering graph (i.e., LightGCN) is highly prone to high-degree nodes and has poor performance on long-tail items ($G_1$-$G_3$).
Other GNN-based models achieve better performance on long-tail items by introducing auxiliary information or constructing self-supervised signals.
Our method L-IMCAT achieves the best performance on long-tail items since it explicitly introduces the item set information to strengthen the supervision signal on items with fewer interactions.

% \subsubsubsection{Cold Start User}
% \subsubsection{long-tail items}

\subsubsection{\textbf{Benefits for Cold-Start Users}}
\begin{figure}[t!]
    \centering
    \centering
    \includegraphics[width=\linewidth]{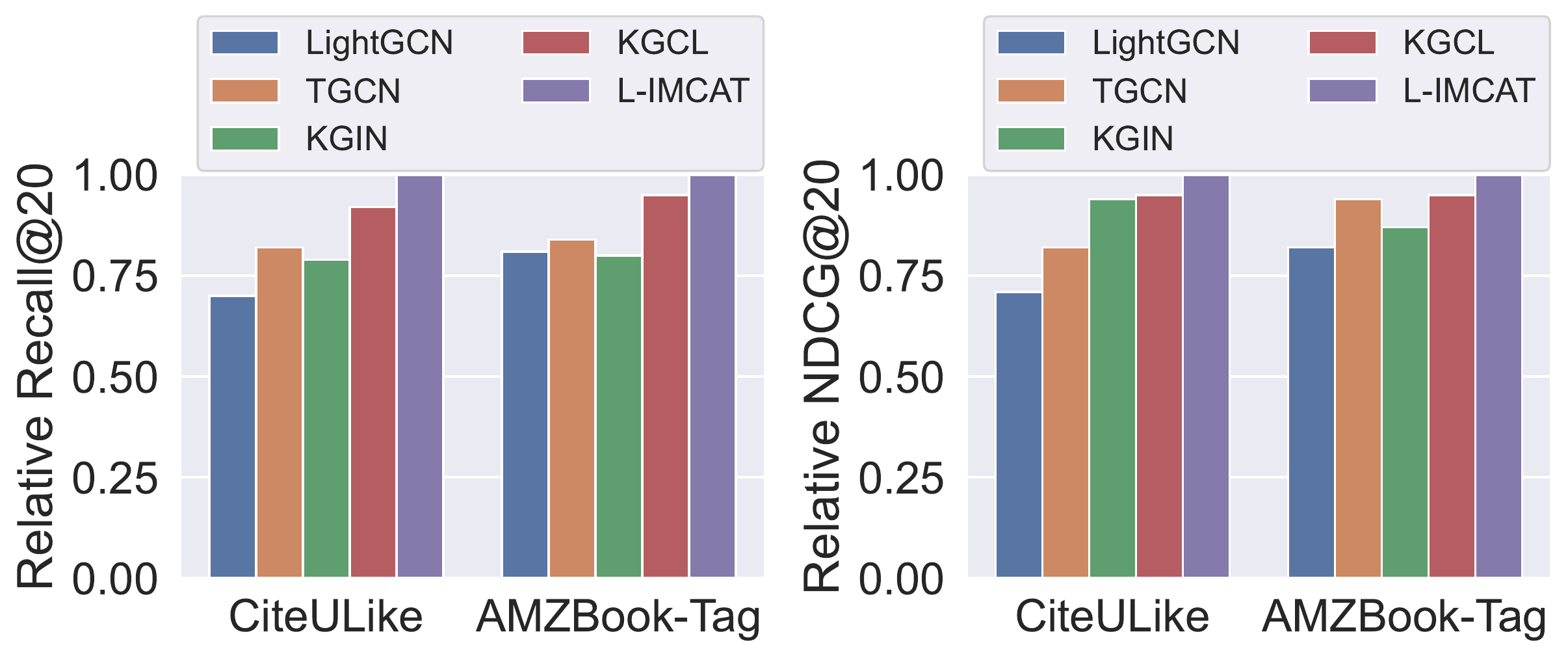}
    \caption{Performance comparison on Cold-start users on different datasets. Performance is normalized into range $[0, 1]$ for each dataset.}
    \label{fig:cold-start}
    \vspace{-5mm}
\end{figure}

We also analyze the performance of our method on cold-start users.
We follow a similar approach to~\cite{YuYLWH021} to generate a sparse user set with fewer than 10 interactions for CiteULike and AMZBook-Tag. 
Similar to our previous experiment, we choose the same GNN-based models for comparison and normalize the performance on each dataset into the range [0, 1] by dividing by the performance obtained by the best model on that dataset.
We report the results in Fig.~\ref{fig:cold-start}.
As illustrated, our proposed method can achieve a good performance on users with fewer interactions compared to prior approaches. 
% \MJC{Any comments on Yelp-Tag? - why does KGCL outperform?}
This indicates that the contrastive alignment among multiple sources can introduce more useful supervision signals not only for items with low degrees but also for cold start users.

\subsubsection{\textbf{Training Efficiency vs Recommendation Quality}}
\begin{figure}[t!]
    \centering
    \centering
    \includegraphics[width=\linewidth]{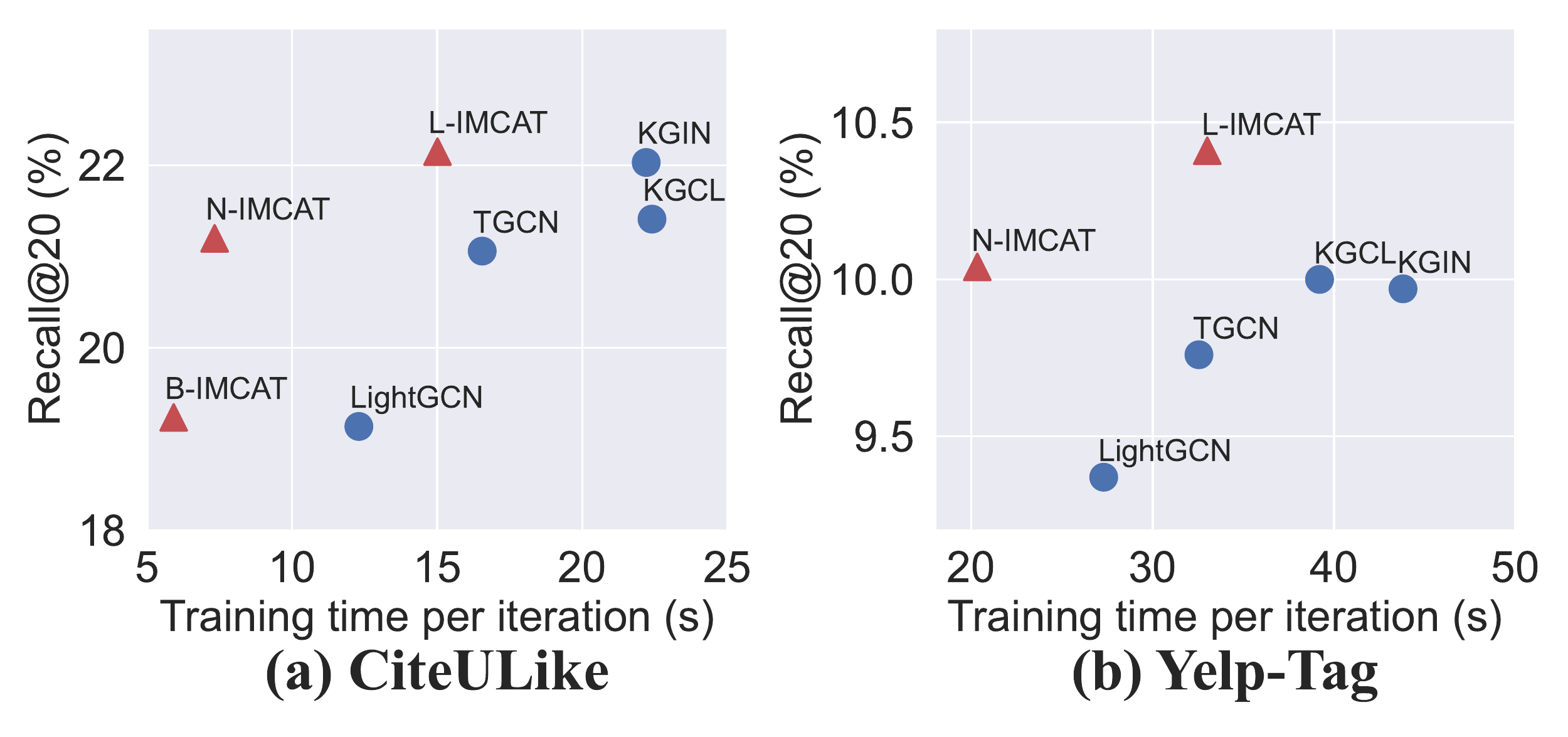}
    \caption{The analysis on the training efficiency versus the recommendation quality.}
    \label{fig:time}
    \vspace{-7mm}
\end{figure}

We investigate the trade-off between training efficiency and  model performance on two datasets. 
As shown in Fig.~\ref{fig:time}, we observe that our proposed methods can achieve a better recommendation performance with reduced training time requirements compared to prior SOTA methods.
For example, a simple NeuMF with our designed IMCAT strategy (i.e., N-IMCAT) can achieve competitive performance with SOTA GNN-based methods (e.g., KGCL) but with more than 50\% training time reduction.
\new{We think the reasons are that GNN-based methods require (a) multi-layer message passing and (b) multi-layer neighborhood sampling for each node, which both lead to more times of computation.
However, both operations are not required in our method.}
This indicates that our proposed intent-aware multi-source contrastive alignment is more efficient than a GNN in this tag-enhanced recommendation task.
This also suggests the potential of applying our proposed method on real-world web-scale datasets.

% \subsubsection{\textbf{Case Study}}

\section{Conclusion}
In this work, we propose a novel tag-enhanced recommendation method for efficiently connecting the collaborative filtering signal and auxiliary tag information from multiple sources (i.e., users, items, and tags) through self-supervised learning.
We construct the correct pairing between the representations obtained from the users and tags that are associated with each item, and then aim to maximize the alignment of each pair's embeddings.
This process is called as multi-source alignment.
Considering there exist many factors behind the user behavior, we also capture user intents in our alignment procedure by decomposing user and item representations into several components.
Each component represents an intent, which is also related to a cluster of tags.
Experiments on three recommendation backbones and seven datasets prove the flexibility and compatibility of our method.
Further analysis demonstrates its superior training efficiency, which strongly suggests the possibility of applying IMCAT on large-scale industrial datasets.

% \begin{table}[t]
% \caption{Training time per epoch.}
%     \centering
%     \begin{tabular}{c|c|c|c|c}
%     \toprule
%          Model &  CiteULike & Last.fm & Amazon\_Book & Yelp2018\\
%          \midrule
%         KGAT & 67.2 & 195.6 & 250.3 & 268.7\\
%         N-IMCAT & 17.9 & 52.1 & 80.8 & 91.2\\
%         \bottomrule
%     \end{tabular}
%     \label{tab:my_label}
% \end{table}

% \begin{table}[t]
% \caption{Caption}
%     \centering
%     \begin{tabular}{c|c|c|c|c}
%     \toprule
%          RAM &  CiteULike & Last.fm & Amazon\_Book & Yelp2018\\
%          \midrule
%         KGAT & 67.2 & 195.6 & 250.3 & 268.7\\
%         BPRMF-IMCA & 17.9 & 52.1 & 80.8 & 91.2\\
%         \bottomrule
%     \end{tabular}
%     \label{tab:my_label}
% \end{table}

% \begin{table}[t]
% \caption{Epochs for reaching 95\% KGAT best performance.}
%     \centering
%     \begin{tabular}{c|c|c|c|c}
%     \toprule
%          Epoch &  CiteULike & Last.fm & Amazon\_Book & Yelp2018\\
%          \midrule
%         KGAT & 300 & 350 & 850 & 650\\
%         N-IMCAT & 150 & 150 & 300 & 350\\
%         \bottomrule
%     \end{tabular}
%     \label{tab:my_label}
% \end{table}

% \subsection{Ablation Analysis on Joint Training}
% UIT (-it), UIT (-ui), UIT (-uit)

% \subsection{Ablation Analysis on Alignment Design}
% Compare the first two alignment

% \subsection{Ablation Analysis on end-to-end clustering}
% MF-UIT, MF-UIT+kmeans, MF-UIT+self

% \subsection{Ablation Analysis on Soft alignment}
% MF-UIT, MF-UIT-AES+hard, MF-UIT-AES+soft

% \section*{Acknowledgement}
% We sincerely appreciate the constructive suggestions from anonymous reviewers. 

\newpage
\bibliographystyle{IEEEtran}
\bibliography{IEEEabrv,references.bib}

\begin{thebibliography}{10}
\providecommand{\url}[1]{#1}
\csname url@samestyle\endcsname
\providecommand{\newblock}{\relax}
\providecommand{\bibinfo}[2]{#2}
\providecommand{\BIBentrySTDinterwordspacing}{\spaceskip=0pt\relax}
\providecommand{\BIBentryALTinterwordstretchfactor}{4}
\providecommand{\BIBentryALTinterwordspacing}{\spaceskip=\fontdimen2\font plus
\BIBentryALTinterwordstretchfactor\fontdimen3\font minus
  \fontdimen4\font\relax}
\providecommand{\BIBforeignlanguage}[2]{{%
\expandafter\ifx\csname l@#1\endcsname\relax
\typeout{** WARNING: IEEEtran.bst: No hyphenation pattern has been}%
\typeout{** loaded for the language `#1'. Using the pattern for}%
\typeout{** the default language instead.}%
\else
\language=\csname l@#1\endcsname
\fi
#2}}
\providecommand{\BIBdecl}{\relax}
\BIBdecl

\bibitem{heuristic_CantadorBV10}
I.~Cantador, A.~Bellog{\'{\i}}n, and D.~Vallet, ``Content-based recommendation
  in social tagging systems,'' in \emph{ACM Recommender Systems Conf.}, 2010.

\bibitem{Linear_Fernandez-TobiasC14}
I.~Fern{\'{a}}ndez{-}Tob{\'{\i}}as and I.~Cantador, ``Exploiting social tags in
  matrix factorization models for cross-domain collaborative filtering,'' in
  \emph{{CEUR} Workshop Proceedings}, 2014.

\bibitem{Rendle10_FactorizationMachine}
S.~Rendle, ``Factorization machines,'' in \emph{{IEEE Int. Conf. on Data
  Mining}}, 2010.

\bibitem{CFA_ZuoZGJ16}
Y.~Zuo, J.~Zeng, M.~Gong, and L.~Jiao, ``Tag-aware recommender systems based on
  deep neural networks,'' \emph{Neurocomputing}, 2016.

\bibitem{DSPR_XuCLMM16}
Z.~Xu, C.~Chen, T.~Lukasiewicz, Y.~Miao, and X.~Meng, ``Tag-aware personalized
  recommendation using a deep-semantic similarity model with negative
  sampling,'' in \emph{{Proc. Conf. on Information and Knowledge Management}},
  2016.

\bibitem{WangZWZLXG18_RippleNet}
H.~Wang, F.~Zhang, J.~Wang, M.~Zhao, W.~Li, X.~Xie, and M.~Guo, ``Ripplenet:
  Propagating user preferences on the knowledge graph for recommender
  systems,'' in \emph{{Proc. Conf. on Information and Knowledge Management}},
  2018.

\bibitem{WangZXLG19_KGCN}
H.~Wang, M.~Zhao, X.~Xie, W.~Li, and M.~Guo, ``Knowledge graph convolutional
  networks for recommender systems,'' in \emph{{Proc. Int. Conf. World Wide
  Web}}, 2019.

\bibitem{KGAT_Wang00LC19}
X.~Wang, X.~He, Y.~Cao, M.~Liu, and T.~Chua, ``{KGAT:} knowledge graph
  attention network for recommendation,'' in \emph{{Proc. Conf. Knowledge
  Discovery and Data Mining}}, 2019.

\bibitem{TGCN_ChenGTXDHW20}
B.~Chen, W.~Guo, R.~Tang, X.~Xin, Y.~Ding, X.~He, and D.~Wang, ``{TGCN:} tag
  graph convolutional network for tag-aware recommendation,'' in \emph{{Proc.
  Conf. on Information and Knowledge Management}}, 2020.

\bibitem{WangJZ0XC20DGCF}
X.~Wang, H.~Jin, A.~Zhang, X.~He, T.~Xu, and T.~Chua, ``Disentangled graph
  collaborative filtering,'' in \emph{{Proc. Conf. on Research and Development
  in Information Retrieval}}, 2020.

\bibitem{ZhangYLXM16CKE}
F.~Zhang, N.~J. Yuan, D.~Lian, X.~Xie, and W.~Ma, ``Collaborative knowledge
  base embedding for recommender systems,'' in \emph{{Proc. Conf. Knowledge
  Discovery and Data Mining}}, 2016.

\bibitem{ClausetSN09_powerlaw}
A.~Clauset, C.~R. Shalizi, and M.~E.~J. Newman, ``Power-law distributions in
  empirical data,'' \emph{{SIAM} Rev.}, 2009.

\bibitem{Anderson_LongTail}
C.~Anderson, \emph{The Long Tail: Why the Future of Business Is Selling Less of
  More}.\hskip 1em plus 0.5em minus 0.4em\relax Hyperion, 2006.

\bibitem{RadfordKHRGASAM21_CLIP}
A.~Radford, J.~W. Kim, C.~Hallacy, A.~Ramesh, G.~Goh, S.~Agarwal, G.~Sastry,
  A.~Askell, P.~Mishkin, J.~Clark, G.~Krueger, and I.~Sutskever, ``Learning
  transferable visual models from natural language supervision,'' in
  \emph{{Proc. Int. Conf. on Machine Learning}}, 2021.

\bibitem{yao2021filip}
L.~Yao, R.~Huang, L.~Hou, G.~Lu, M.~Niu, H.~Xu, X.~Liang, Z.~Li, X.~Jiang, and
  C.~Xu, ``Filip: Fine-grained interactive language-image pre-training,''
  \emph{arXiv preprint arXiv:2111.07783}, 2021.

\bibitem{HjelmFLGBTB19_MIE}
R.~D. Hjelm, A.~Fedorov, S.~Lavoie{-}Marchildon, K.~Grewal, P.~Bachman,
  A.~Trischler, and Y.~Bengio, ``Learning deep representations by mutual
  information estimation and maximization,'' in \emph{{Proc. Int. Conf. on
  Learning Representations}}, 2019.

\bibitem{LarssonMS16_color}
G.~Larsson, M.~Maire, and G.~Shakhnarovich, ``Learning representations for
  automatic colorization,'' in \emph{{European Conf. on Computer Vision}}, ser.
  Lecture Notes in Computer Science, 2016.

\bibitem{abs-1807-03748_coding}
A.~Oord, Y.~Li, and O.~Vinyals, ``Representation learning with contrastive
  predictive coding,'' \emph{CoRR}, vol. abs/1807.03748, 2018.

\bibitem{RendleFGS2009_bpr}
S.~Rendle, C.~Freudenthaler, Z.~Gantner, and L.~Schmidt{-}Thieme, ``{BPR:}
  bayesian personalized ranking from implicit feedback,'' in \emph{{Proc. Conf.
  on Uncertainty in Artificial Intelligence}}, 2009.

\bibitem{JIT2R}
X.~Chen, C.~Du, X.~He, and J.~Wang, ``{JIT2R:} {A} joint framework for item
  tagging and tag-based recommendation,'' in \emph{{Proc. Conf. on Research and
  Development in Information Retrieval}}, 2020.

\bibitem{HDLPR_XuLCMM17}
Z.~Xu, T.~Lukasiewicz, C.~Chen, Y.~Miao, and X.~Meng, ``Tag-aware personalized
  recommendation using a hybrid deep model,'' in \emph{{Proc. Int. Joint Conf.
  on Artificial Intelligence}}, 2017.

\bibitem{0002HLL18KGIntro1}
Y.~Cao, L.~Hou, J.~Li, and Z.~Liu, ``Neural collective entity linking,'' in
  \emph{{Int. Conf. on Computational Linguistics}}, 2018.

\bibitem{0002HLLLCD18KGINtro2}
Y.~Cao, L.~Hou, J.~Li, Z.~Liu, C.~Li, X.~Chen, and T.~Dong, ``Joint
  representation learning of cross-lingual words and entities via attentive
  distant supervision,'' in \emph{{Empirical Methods in Natural Language
  Processing}}, 2018.

\bibitem{YuRSGSKNH14PER}
X.~Yu, X.~Ren, Y.~Sun, Q.~Gu, B.~Sturt, U.~Khandelwal, B.~Norick, and J.~Han,
  ``Personalized entity recommendation: a heterogeneous information network
  approach,'' in \emph{{Proc. Conf. Web Search and Data Mining}}, 2014.

\bibitem{HuSZY18MCRec}
B.~Hu, C.~Shi, W.~X. Zhao, and P.~S. Yu, ``Leveraging meta-path based context
  for top- {N} recommendation with {A} neural co-attention model,'' in
  \emph{{Proc. Conf. Knowledge Discovery and Data Mining}}, 2018.

\bibitem{XiaHXDZYPB21KGBehavior}
L.~Xia, C.~Huang, Y.~Xu, P.~Dai, X.~Zhang, H.~Yang, J.~Pei, and L.~Bo,
  ``Knowledge-enhanced hierarchical graph transformer network for
  multi-behavior recommendation,'' in \emph{{Proc. Conf. Uncertainty in
  Artificial Intelligence}}, 2021.

\bibitem{WangZXG18DKN}
H.~Wang, F.~Zhang, X.~Xie, and M.~Guo, ``{DKN:} deep knowledge-aware network
  for news recommendation,'' in \emph{{Proc. Int. Conf. World Wide Web}}, 2018.

\bibitem{Xin0ZZJ19RCF}
X.~Xin, X.~He, Y.~Zhang, Y.~Zhang, and J.~M. Jose, ``Relational collaborative
  filtering: Modeling multiple item relations for recommendation,'' in
  \emph{{Proc. Conf. on Research and Development in Information Retrieval}},
  2019.

\bibitem{TianYRWWWL21KGNews}
Y.~Tian, Y.~Yang, X.~Ren, P.~Wang, F.~Wu, Q.~Wang, and C.~Li, ``Joint knowledge
  pruning and recurrent graph convolution for news recommendation,'' in
  \emph{{Proc. Conf. on Research and Development in Information Retrieval}},
  2021.

\bibitem{LinLSLZ15TransR}
Y.~Lin, Z.~Liu, M.~Sun, Y.~Liu, and X.~Zhu, ``Learning entity and relation
  embeddings for knowledge graph completion,'' in \emph{{Proc. Conf.
  Uncertainty in Artificial Intelligence}}, 2015.

\bibitem{WangHWYL0C21_KGIN}
X.~Wang, T.~Huang, D.~Wang, Y.~Yuan, Z.~Liu, X.~He, and T.~Chua, ``Learning
  intents behind interactions with knowledge graph for recommendation,'' in
  \emph{{Proc. Int. Conf. World Wide Web}}, 2021.

\bibitem{ChenK0H20_SimCLR}
T.~Chen, S.~Kornblith, M.~Norouzi, and G.~E. Hinton, ``A simple framework for
  contrastive learning of visual representations,'' in \emph{{Proc. Int. Conf.
  on Machine Learning}}, 2020.

\bibitem{HjelmFLGBTB19_MI}
R.~D. Hjelm, A.~Fedorov, S.~Lavoie{-}Marchildon, K.~Grewal, P.~Bachman,
  A.~Trischler, and Y.~Bengio, ``Learning deep representations by mutual
  information estimation and maximization,'' in \emph{{Proc. Int. Conf. on
  Learning Representations}}, 2019.

\bibitem{He0WXG20_momentum}
K.~He, H.~Fan, Y.~Wu, S.~Xie, and R.~B. Girshick, ``Momentum contrast for
  unsupervised visual representation learning,'' in \emph{{Conf. on Computer
  Vision and Pattern Recognition}}, 2020.

\bibitem{GrillSATRBDPGAP20_BYOL}
J.~Grill, F.~Strub, F.~Altch{\'{e}}, C.~Tallec, P.~H. Richemond,
  E.~Buchatskaya, C.~Doersch, B.~{\'{A}}. Pires, Z.~Guo, M.~G. Azar, B.~Piot,
  K.~Kavukcuoglu, R.~Munos, and M.~Valko, ``Bootstrap your own latent - {A} new
  approach to self-supervised learning,'' in \emph{Proc. Adv. Neural Inf. Proc.
  Systems}, 2020.

\bibitem{ChenH21_SimSiam}
X.~Chen and K.~He, ``Exploring simple siamese representation learning,'' in
  \emph{{Conf. on Computer Vision and Pattern Recognition}}, 2021.

\bibitem{S3Rec}
K.~Zhou, H.~Wang, W.~X. Zhao, Y.~Zhu, S.~Wang, F.~Zhang, Z.~Wang, and J.~Wen,
  ``S{\^{}}3-rec: Self-supervised learning for sequential recommendation with
  mutual information maximization,'' \emph{CoRR}, vol. abs/2008.07873, 2020.

\bibitem{XinKAJ20_SQN}
X.~Xin, A.~Karatzoglou, I.~Arapakis, and J.~M. Jose, ``Self-supervised
  reinforcement learning for recommender systems,'' in \emph{{Proc. Conf. on
  Research and Development in Information Retrieval}}, 2020.

\bibitem{YaoYCYCMHCTKE21_google_SSLRS}
T.~Yao, X.~Yi, D.~Z. Cheng, F.~X. Yu, T.~Chen, A.~K. Menon, L.~Hong, E.~H. Chi,
  S.~Tjoa, J.~J. Kang, and E.~Ettinger, ``Self-supervised learning for
  large-scale item recommendations,'' in \emph{{Proc. Conf. on Information and
  Knowledge Management}}, 2021.

\bibitem{Wu_SGL}
J.~Wu, X.~Wang, F.~Feng, X.~He, L.~Chen, J.~Lian, and X.~Xie, ``Self-supervised
  graph learning for recommendation,'' \emph{CoRR}, vol. abs/2010.10783, 2020.

\bibitem{abs-2205-00976KGCL}
Y.~Yang, C.~Huang, L.~Xia, and C.~Li, ``Knowledge graph contrastive learning
  for recommendation,'' \emph{CoRR}, vol. abs/2205.00976, 2022.

\bibitem{Lloyd1982Kmeans}
S.~P. Lloyd, ``Least squares quantization in pcm,'' in \emph{{Information
  Theory, IEEE Transactions}}, 1982.

\bibitem{WangPHLJZ19AttributeCluster}
C.~Wang, S.~Pan, R.~Hu, G.~Long, J.~Jiang, and C.~Zhang, ``Attributed graph
  clustering: {A} deep attentional embedding approach,'' in \emph{{Proc. Int.
  Joint Conf. on Artificial Intelligence}}, 2019.

\bibitem{Bo0SZL020StructureCluster}
D.~Bo, X.~Wang, C.~Shi, M.~Zhu, E.~Lu, and P.~Cui, ``Structural deep clustering
  network,'' in \emph{{Proc. Int. Conf. World Wide Web}}, 2020.

\bibitem{Maaten_student}
L.~van~der Maaten and G.~E. Hinton, ``Visualizing data using t-sne,''
  \emph{Journal of Machine Learning Research}.

\bibitem{ChenK0H20SimCLR}
T.~Chen, S.~Kornblith, M.~Norouzi, and G.~E. Hinton, ``A simple framework for
  contrastive learning of visual representations,'' in \emph{{Proc. Int. Conf.
  on Machine Learning}}, 2020.

\bibitem{oord2018representation}
A.~Oord, Y.~Li, and O.~Vinyals, ``Representation learning with contrastive
  predictive coding,'' \emph{arXiv preprint arXiv:1807.03748}, 2018.

\bibitem{HetRec}
\emph{HetRec '11: Proceedings of the 2nd International Workshop on Information
  Heterogeneity and Fusion in Recommender Systems}.\hskip 1em plus 0.5em minus
  0.4em\relax Association for Computing Machinery, 2011.

\bibitem{WangCL13CiteULike}
H.~Wang, B.~Chen, and W.~Li, ``Collaborative topic regression with social
  regularization for tag recommendation,'' in \emph{{Proc. Int. Joint Conf. on
  Artificial Intelligence}}, 2013.

\bibitem{HeM16AmazonData}
R.~He and J.~J. McAuley, ``Ups and downs: Modeling the visual evolution of
  fashion trends with one-class collaborative filtering,'' in \emph{{Proc. Int.
  Conf. World Wide Web}}, 2016.

\bibitem{sun2020_mgcf}
J.~Sun, Y.~Zhang, C.~Ma, M.~Coates, H.~Guo, R.~Tang, and X.~He, ``Multi-graph
  convolution collaborative filtering,'' in \emph{IEEE Int. Conf. on Data
  Mining}, 2019.

\bibitem{DBLP:conf/wsdm/WangF0NC21}
W.~Wang, F.~Feng, X.~He, L.~Nie, and T.~Chua, ``Denoising implicit feedback for
  recommendation,'' in \emph{{Proc. Conf. Web Search and Data Mining}}, 2021.

\bibitem{Ma2021_hyperknow}
C.~Ma, L.~Ma, Y.~Zhang, H.~Wu, X.~Liu, and M.~Coates, ``Knowledge-enhanced
  top-k recommendation in poincar{\'{e}} ball,'' in \emph{{Proc. Conf.
  Uncertainty in Artificial Intelligence}}, 2021.

\bibitem{JarvelinK02ndcg}
K.~J{\"{a}}rvelin and J.~Kek{\"{a}}l{\"{a}}inen, ``Cumulated gain-based
  evaluation of {IR} techniques,'' \emph{{ACM} Trans. Inf. Syst.}, 2002.

\bibitem{KorenBV09MF}
Y.~Koren, R.~M. Bell, and C.~Volinsky, ``Matrix factorization techniques for
  recommender systems,'' \emph{Computer}, 2009.

\bibitem{HeLZNHC17NeuMF}
X.~He, L.~Liao, H.~Zhang, L.~Nie, X.~Hu, and T.~Chua, ``Neural collaborative
  filtering,'' in \emph{{Proc. Int. Conf. World Wide Web}}, 2017.

\bibitem{he2020_lgcn}
X.~He, K.~Deng, X.~Wang, Y.~Li, Y.~Zhang, and M.~Wang, ``Lightgcn: Simplifying
  and powering graph convolution network for recommendation,'' in \emph{{Proc.
  Conf. on Research and Development in Information Retrieval}}, 2020.

\bibitem{GlorotB10Xavier_init}
X.~Glorot and Y.~Bengio, ``Understanding the difficulty of training deep
  feedforward neural networks,'' in \emph{{Proc. Int. Conf. on Artificial
  Intelligence and Statistics}}, 2010.

\bibitem{YuYLWH021}
J.~Yu, H.~Yin, J.~Li, Q.~Wang, N.~Q.~V. Hung, and X.~Zhang, ``Self-supervised
  multi-channel hypergraph convolutional network for social recommendation,''
  in \emph{{Proc. Int. Conf. World Wide Web}}, 2021.

\end{thebibliography}
% \bibliographystyle{ACM-Reference-Format.bst}
% \bibliography{references.bib}

\end{document}